\documentclass[11pt]{article}
\usepackage{latexsym}
\usepackage{verbatim}
\usepackage[usenames]{color}
\usepackage{graphicx}
\usepackage{amsfonts}
\usepackage{here}
\makeatletter

\@addtoreset{equation}{section}

\@addtoreset{table}{section}

\@addtoreset{figure}{section}
\def\section{\@startsection{section}{1}{0pt}{-3.5ex plus -1ex minus
 -.2ex}{2.3ex plus .2ex}{\large\bf}}
\def\subsection{\@startsection{subsection}{2}{\z@}{-3.25ex plus -1ex minus
 -.2ex}{1.5ex plus .2ex}{\normalsize\bf}}
\def\subsubsection{\@startsection{subsubsection}{3}{\z@}{-3.25ex plus
 -1ex minus -.2ex}{1.5ex plus .2ex}{\normalsize\bf}}
\makeatother
\begin{document}
\newcommand{\bm}{\boldmath}
\newcommand{\Frac}[2]{\displaystyle\frac{#1}{#2}}
\newcommand{\delfrac}[2]{\displaystyle\frac{\partial #1}{\partial #2}}
\newcommand{\dfrac}[2]{\displaystyle\frac{d #1}{d #2}}
\newcommand{\Int}{\displaystyle{\int}}
\newcommand{\iprod}{\, \mbox{\raise.3ex\hbox{\tiny $\bullet$}}\, }
\newcommand{\cH}{\stackrel{\circ}{H}}
\newcommand{\sgn}{\mbox{sgn }}
\newcommand{\sdim}{d}
\newcommand{\Backslash}[1]{\ooalign{\hfil$\backslash$\hfil\crcr$#1$}}
\newcommand{\diffrac}[2]{\displaystyle\frac{d #1}{d #2}}
\newcommand{\QED}{$\Box$} 
\newtheorem{definition}{\textcolor{blue}{Definition}}[section]
\newtheorem{assumption}{\textcolor{blue}{Assumption}}[section]
\newtheorem{theorem}{\textcolor{blue}{Theorem}}[section]
\newtheorem{lemma}{\textcolor{blue}{Lemma}}[section]
\newtheorem{remark}{\textcolor{blue}{Remark}}[section]
\newtheorem{algorithm}{\textcolor{blue}{Algorithm}}[section]

\begin{center} 
\Large{Finding Similar Objects and Active Inference
for Surprise} 
\\ 
\Large{in Numenta Neocortex Model} 
\end{center} 
\vspace{5pt} 

\begin{center} 
Hajime Kawakami
\\
Akita University, Akita, 010-8502, Japan
\\
kawakami@math.akita-u.ac.jp, 
\hspace{5pt}
hjm.kwkm.07091210@gmail.com
\end{center}
\vspace{5pt} 
 
\begin{center} 
June 11, 2025
\end{center} 
\vspace{10pt} 

\section*{Abstract}

Jeff Hawkins and his colleagues in Numenta have proposed
the thousand-brains system.
This is a model of the structure and operation of the neocortex
and is under investigation as a new form of artificial intelligence.
In their study, learning and inference algorithms 
running on the system are proposed,
where the prediction is an important function.
The author believes that one of the most important capabilities 
of the neocortex 
in addition to prediction
is the ability to make association,
that is, to find the relationships between objects.
Similarity is an important example of such relationships.
In our study, algorithms 
that run on the thousand-brains system 
to find similarities are proposed.
Although the setting for 
these algorithms is restricted, 
the author believes that 
the case it covers is fundamental. 
Karl Friston and his colleagues have studied the free-energy principle  
that explains how the brain actively infers the cause 
of a Shannon surprise.
In our study, 
an algorithm is proposed for the thousand-brains system
to make this inference.
The problem of inferring what is being observed from the sensory data is 
a type of inverse problem, and
the inference algorithms of 
the thousand-brains system and 
free-energy principle 
solve this problem in a Bayesian manner.
Our inference 
algorithms can also be interpreted
as Bayesian or non-Bayesian updating processes.
\vspace{10pt} 

\noindent
\textbf{Keywords}\hspace{5pt}  
Neocortex $\cdot$ Thousand Brains $\cdot$ 
Similarity $\cdot$ Active Inference $\cdot$ Bayesian Inference   
$\cdot$ non-Bayesian Inference $\cdot$ Inverse Problem

\section{Introduction}
\label{sect.introduction}

Conventionally, scientists state that the neocortex of the brain 
vertically comprises six layers. Thus, the layers run parallel to the surface 
of the neocortex. 
The neocortex is horizontally divided into several 
regions such as the visual and touch regions. 
For instance, the visual region comprises  
several areas such as V1, V2, and V3.
The neocortex, each region, and each area  
comprise numerous cortical columns that 
penetrate the six layers. 
Numerous feedforward and feedback connections exist between neurons in 
these cortical columns. 

On pages 24 and 25 of \cite{H2}, citing \cite{M}, 
Hawkins states:  
\begin{center}
\begin{minipage}{400pt}
\textit{Mountcastle is proposing that all the things we associate 
with intelligence, which on the surface appear to be different,
are, in reality, manifestations of the same underlying cortical 
algorithm. $\ldots$ 
So, what was Mountcastle's proposal for the location of the 
cortical algorithm ? He said that the fundamental unit of the 
neocortex, the unit of intelligence, was a ``cortical column.''}
\end{minipage}
\end{center}
However, Mountcastle did not propose any algorithm:
how a cortical column does all the things we associate with 
intelligence.  
Thus, Hawkins et al. in Numenta proposed such algorithms 
in \cite{H1}, \cite{H2}, \cite{HAC}, \cite{HLKPA}, and \cite{LPAH}. 
We refer to these algorithms collectively as 
the \textit{Numenta (neocortex) model}. 
In these studies, prediction is considered as the 
most important capability of the neocortex, and 
algorithms in the cortical columns 
for learning and inference, including prediction, 
have been proposed 
(see Algorithms \ref{algorithm.subsect.model.learning.1} 
and \ref{algorithm.subsect.model.inference.1}
described below). 
The cortical columns learn the structure of objects using this   
learning algorithm, and infer 
the object under observation 
using this inference algorithm with the sensory input.
Hawkins named the system they created, which included 
the Numenta model, the thousand-brains system.  
While writing this manuscript, the paper \cite{CLH} 
by Hawkins et al. was published.  
In this paper, Monty, the first instantiation of 
the thousand-brains system, is proposed.  
Our study is based primarily on \cite{HAC} and \cite{LPAH}, 
which explicitly describe the Numenta model algorithms, 
and it also refers to \cite{CLH}.

What are the other important capabilities 
of the neocortex in addition to prediction ?  
Section 2.4 of \cite{CLH} lists the expected functions 
of a model of the neocortex. Related to this list, the author believes that 
one of the important capabilities is making ``association,'' that is, 
finding the relationships between objects.
Similarity between objects is an important example of such relationships.
The importance of ``association'' has been highlighted 
by numerous scientists. For instance, 
Polya \cite{P} states the following 
on the list entitled ``How to Solve it'':  
\begin{center}
\begin{minipage}{400pt}
\textit{Find the connection between the data and the unknown.}
\end{minipage}
\end{center}
and
\begin{center}
\begin{minipage}{400pt}
\textit{Have you seen it before ? Or have you seen the same problem in a
slightly different form ?}  
\end{minipage}
\end{center}
P. A. M. Dirac states that: 
\begin{center}
\begin{minipage}{400pt}
\textit{With the mathematical procedure there are two main methods that
  one may follow, (i) to remove inconsistencies and (ii) to
  unite theories that were previously disjoint.}
\end{minipage}
\end{center}
on page 58 of \cite{PJO}.  
Hawkins emphasizes the importance of similarities
with respect to Mountcastle's idea (see Chapter 3 of \cite{H1} and 
Chapter 2 of \cite{H2}). 
He also states that:  
\begin{center}
\begin{minipage}{400pt}
\textit{When mathematicians see a new equation, they recognize it 
as similar to previous equations they have worked with.}
\end{minipage}
\end{center}
on page 82 of \cite{H2}. 
In this study, we propose algorithms 
(Algorithms \ref{algorithm.sect.related.1} 
and \ref{algorithm.sect.related.2})  
that run on the Numenta model to find similarities.
These algorithms are based on 
the Numenta inference algorithm  
(Algorithm \ref{algorithm.subsect.model.inference.1}).   
The setting for 
these algorithms is restricted, 
and for more general ``associations,'' this setting  
is significantly limited.
However, 
the author believes that 
the case it covers is fundamental. 

Friston et al. have studied the free-energy principle, 
for instance, \cite{F1}, \cite{F2}, and \cite{PPF}. 
The free-energy principle explains how the brain infers the cause 
for a Shannon surprise (informational surprise).
In our study, 
an algorithm (Algorithm \ref{algorithm.sect.inference.1}) based
on Algorithm \ref{algorithm.subsect.model.inference.1}
is proposed for the Numenta model to obtain
this inference.  
A relationship between the thousand-brains system 
and free-energy principle has been investigated in studies such as  
\cite{VVCD}.
Friston's theory is based on probability theory, 
specifically the Bayesian inference theory.
Inference in the Numenta model is refined 
by reducing the ambiguity based on successive observations. 
Therefore, this inference can be considered to be Bayesian.
The problem of inferring what is being observed from the sensory data is 
a type of inverse problem, and the inference algorithms of 
the Numenta model and free-energy principle 
solve this problem in a Bayesian manner.
We also consider our inference algorithms 
from the perspective of Bayesian inference. 

The remainder of this manuscript is organized as follows: 
In \S \ref{sect.object} and \S \ref{sect.model}, 
the Numenta neocortex model and learning and inference algorithms 
(Algorithms \ref{algorithm.subsect.model.learning.1} 
and \ref{algorithm.subsect.model.inference.1})   
are reviewed.   
However, these algorithms are 
slightly changed, primarily for simplicity. 
In \S \ref{sect.related} and \S \ref{sect.inference},  
by slightly changing the Numenta inference algorithm,
algorithms to find objects that are similar to a given object 
(Algorithms \ref{algorithm.sect.related.1} 
and \ref{algorithm.sect.related.2})  
and an
algorithm to actively infers surprise 
(Algorithm \ref{algorithm.sect.inference.1}) are proposed.

Although the proposed algorithms of 
this study are limited and are not based on brain 
experimental results, 
the author hopes that they will contribute to future studies on 
the brain or 
artificial intelligence. 
Real systems almost always encounter errors, and in the following, 
the equations contain few of such errors, unless otherwise noted. 
It is believed that not only neurons but also 
glial cells are important 
for the transmission of information in the brain
(cf. \cite{F}).  
However, only neurons are considered in the Numenta model and 
this study. 

\section{Object, observation, learning, inference, and recognition} 
\label{sect.object} 

\begin{figure}[H]
\caption{An object and a movement vector} 
\unitlength 1pt
\begin{center} 
\begin{picture}(240,120)
\put(0,110){(A)}
\put(140,110){(B)}
\put(40,30){\line(1,0){40}}
\put(40,50){\line(1,0){60}}
\put(20,70){\line(1,0){80}}
\put(20,90){\line(1,0){80}}
\put(40,110){\line(1,0){20}}
\put(20,70){\line(0,1){20}}
\put(40,30){\line(0,1){80}}
\put(60,30){\line(0,1){80}}
\put(80,30){\line(0,1){60}}
\put(100,50){\line(0,1){40}}
\put(48,38){$\star$} 
\put(68,38){$\circ$} 
\put(47,57){$\Box$} 
\put(68,58){$\circ$} 
\put(88,58){$\circ$} 
\put(28,78){$\star$} 
\put(47,77){$\Box$} 
\put(67,77){$\Box$} 
\put(88,78){$\star$}
\put(48,98){$\star$}
\put(180,30){\line(1,0){40}}
\put(180,50){\line(1,0){60}}
\put(160,70){\line(1,0){80}}
\put(160,90){\line(1,0){80}}
\put(180,110){\line(1,0){20}}
\put(160,70){\line(0,1){20}}
\put(180,30){\line(0,1){80}}
\put(200,30){\line(0,1){80}}
\put(220,30){\line(0,1){60}}
\put(240,50){\line(0,1){40}}
\put(188,38){$\star$} 
\put(208,38){$\circ$} 
\put(187,57){$\Box$} 
\put(208,58){$\circ$} 
\put(228,58){$\circ$} 
\put(168,78){$\star$} 
\put(187,77){$\Box$} 
\put(207,77){$\Box$} 
\put(228,78){$\star$}
\put(188,98){$\star$}
\put(190,40){\textcolor{red}{\vector(1,2){20}}}
\put(60,10){$O$}  
\put(200,10){$O$}
\end{picture}  
\end{center} 
\label{eq.subsect.parameters.example.1}
\end{figure}
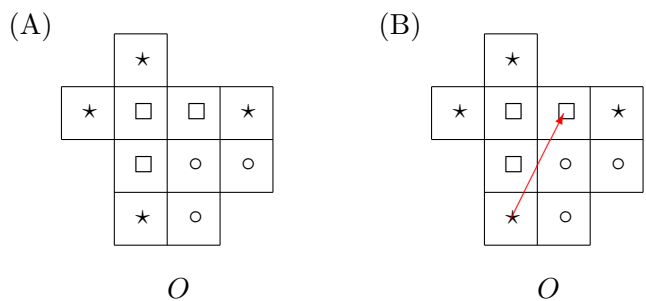

The Numenta model learns, infers, and recognizes objects. 
Figure \ref{eq.subsect.parameters.example.1} (A) 
shows an example of such an object 
(see Figure 2 of \cite{HAC} and Figure 5 of \cite{LPAH}).
This object $O$ comprises ten pairs of locations and features,
(location, feature),
where $\star,$ $\Box,$ and $\circ$ are the features.
(Several cases of more general objects are considered in \cite{CLH}. 
In \S 3.1 of \cite{CLH}, it is stated that 
``\textit{an object is a discrete entity composed of a collection 
of one or more other objects}.'' 
Habitat objects, YCB object dataset,
and other datasets are listed in \S 6 
as objects for simulation.) 
In the Numenta model, time $t$ is a discrete variable,  
$t = 0, 1, 2, \ldots,$ 
and each cortical column of 
the model observes and senses one pair of 
$\mbox{(location, feature)}$ at each time step. 
Figure \ref{eq.subsect.parameters.example.1} (B)  
illustrates an example of an observation of $O.$
The first observation location is the starting point of the red arrow, and 
the sensory feature is $\star$ at this location.  
The next observation location is the end point of the arrow, and 
the sensory feature is $\Box$ at this location.  
Such an arrow is called a movement vector
in \cite{HAC} and \cite{LPAH}. 
Thus, a cortical column of 
the Numenta model learns $O$ by 
observing and sensing 
pairs $\mbox{(location, feature)}$ individually 
(by Algorithm \ref{algorithm.subsect.model.learning.1}). 

As described below, the inference is also performed by 
observing and sensing 
pairs (location, feature) individually 
(by Algorithm \ref{algorithm.subsect.model.inference.1}). 
Assume that 
the model has already learned objects 
$O,$ $O',$ and $O''$ of Figure \ref{eq.sect.object.1.Obj},   
and then begins observing and inferring object $O.$
In Figure \ref{eq.sect.object.1.ABC} (A), 
the red arrow represents the first real movement vector. 
Then, both the first and second sensory features are $\circ,$
and this movement can not be distinguished from the other movements 

\begin{figure}[H]
\caption{Example of objects} 
\unitlength 1pt
\begin{center}
\begin{picture}(270,120)
\put(20,30){\line(1,0){40}}
\put(20,50){\line(1,0){60}}
\put(0,70){\line(1,0){80}}
\put(0,90){\line(1,0){80}}
\put(20,110){\line(1,0){20}}
\put(0,70){\line(0,1){20}}
\put(20,30){\line(0,1){80}}
\put(40,30){\line(0,1){80}}
\put(60,30){\line(0,1){60}}
\put(80,50){\line(0,1){40}}
\put(28,38){$\star$} 
\put(48,38){$\circ$} 
\put(27,57){$\Box$} 
\put(48,58){$\circ$} 
\put(68,58){$\circ$} 
\put(8,78){$\star$} 
\put(27,77){$\Box$} 
\put(47,77){$\Box$} 
\put(68,78){$\star$}
\put(28,98){$\star$}
\put(140,30){\line(1,0){20}}
\put(120,50){\line(1,0){60}}
\put(100,70){\line(1,0){80}}
\put(100,90){\line(1,0){80}}
\put(120,110){\line(1,0){20}}
\put(100,70){\line(0,1){20}}
\put(120,50){\line(0,1){60}}
\put(140,30){\line(0,1){80}}
\put(160,30){\line(0,1){60}}
\put(180,50){\line(0,1){40}}
\put(147,37){$\Box$} 
\put(128,58){$\circ$} 
\put(147,57){$\Box$} 
\put(167,57){$\Box$} 
\put(108,78){$\star$} 
\put(128,78){$\circ$} 
\put(148,78){$\circ$} 
\put(168,78){$\star$}
\put(128,98){$\star$} 
\put(220,30){\line(1,0){40}}
\put(220,50){\line(1,0){60}}
\put(200,70){\line(1,0){80}}
\put(200,90){\line(1,0){80}}
\put(260,110){\line(1,0){20}}
\put(200,70){\line(0,1){20}}
\put(220,30){\line(0,1){60}}
\put(240,30){\line(0,1){60}}
\put(260,30){\line(0,1){80}}
\put(280,50){\line(0,1){60}}
\put(228,38){$\bullet$} 
\put(248,38){$\circ$} 
\put(227,57){$\Box$} 
\put(248,58){$\circ$} 
\put(267,57){$\Box$} 
\put(208,78){$\bullet$} 
\put(227,77){$\circ$} 
\put(247,77){$\Box$} 
\put(268,78){$\bullet$}
\put(268,98){$\bullet$}
\put(40,10){$O$}  
\put(140,10){$O'$}
\put(240,10){$O''$}    
\end{picture} 
\end{center}
\label{eq.sect.object.1.Obj} 
\end{figure}
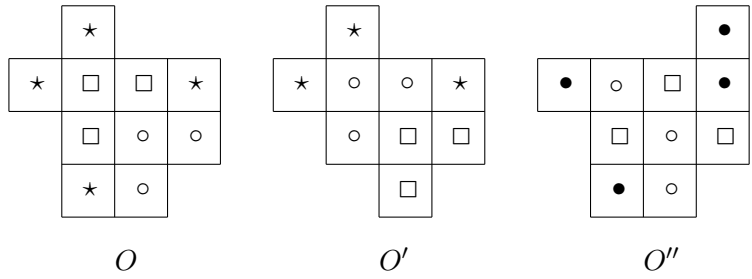

\begin{figure}[H]
\caption{Convergence onto a representation for $O$} 
\unitlength 1pt
\begin{center}
\begin{picture}(290,130)
\put(-20,110){(A)} 
\put(20,30){\line(1,0){40}}
\put(20,50){\line(1,0){60}}
\put(0,70){\line(1,0){80}}
\put(0,90){\line(1,0){80}}
\put(20,110){\line(1,0){20}}
\put(0,70){\line(0,1){20}}
\put(20,30){\line(0,1){80}}
\put(40,30){\line(0,1){80}}
\put(60,30){\line(0,1){60}}
\put(80,50){\line(0,1){40}}
\put(28,38){$\star$} 
\put(48,38){$\circ$} 
\put(27,57){$\Box$} 
\put(48,58){$\circ$} 
\put(68,58){$\circ$} 
\put(8,78){$\star$} 
\put(27,77){$\Box$} 
\put(47,77){$\Box$} 
\put(68,78){$\star$}
\put(28,98){$\star$}
\put(50,63){\textcolor{red}{\vector(1,0){20}}}
\put(70,59){\textcolor{blue}{\vector(-1,0){20}}}
\put(48,58){\textcolor{blue}{\vector(0,-1){20}}}
\put(52,38){\textcolor{blue}{\vector(0,1){20}}}
\put(140,30){\line(1,0){20}}
\put(120,50){\line(1,0){60}}
\put(100,70){\line(1,0){80}}
\put(100,90){\line(1,0){80}}
\put(120,110){\line(1,0){20}}
\put(100,70){\line(0,1){20}}
\put(120,50){\line(0,1){60}}
\put(140,30){\line(0,1){80}}
\put(160,30){\line(0,1){60}}
\put(180,50){\line(0,1){40}}
\put(147,37){$\Box$} 
\put(128,58){$\circ$} 
\put(147,57){$\Box$} 
\put(167,57){$\Box$} 
\put(108,78){$\star$} 
\put(128,78){$\circ$} 
\put(148,78){$\circ$} 
\put(168,78){$\star$}
\put(128,98){$\star$} 
\put(130,83){\textcolor{blue}{\vector(1,0){20}}}
\put(150,79){\textcolor{blue}{\vector(-1,0){20}}}
\put(128,78){\textcolor{blue}{\vector(0,-1){20}}}
\put(132,58){\textcolor{blue}{\vector(0,1){20}}}
\put(220,30){\line(1,0){40}}
\put(220,50){\line(1,0){60}}
\put(200,70){\line(1,0){80}}
\put(200,90){\line(1,0){80}}
\put(260,110){\line(1,0){20}}
\put(200,70){\line(0,1){20}}
\put(220,30){\line(0,1){60}}
\put(240,30){\line(0,1){60}}
\put(260,30){\line(0,1){80}}
\put(280,50){\line(0,1){60}}
\put(228,38){$\bullet$} 
\put(248,38){$\circ$} 
\put(227,57){$\Box$} 
\put(248,58){$\circ$} 
\put(267,57){$\Box$} 
\put(208,78){$\bullet$} 
\put(227,77){$\circ$} 
\put(247,77){$\Box$} 
\put(268,78){$\bullet$}
\put(268,98){$\bullet$}
\put(40,10){$O$}  
\put(140,10){$O'$}
\put(240,10){$O''$}    
\put(248,58){\textcolor{blue}{\vector(0,-1){20}}}
\put(252,38){\textcolor{blue}{\vector(0,1){20}}}
\end{picture}  
\begin{picture}(290,130)
\put(-20,110){(B)} 
\put(20,30){\line(1,0){40}}
\put(20,50){\line(1,0){60}}
\put(0,70){\line(1,0){80}}
\put(0,90){\line(1,0){80}}
\put(20,110){\line(1,0){20}}
\put(0,70){\line(0,1){20}}
\put(20,30){\line(0,1){80}}
\put(40,30){\line(0,1){80}}
\put(60,30){\line(0,1){60}}
\put(80,50){\line(0,1){40}}
\put(28,38){$\star$} 
\put(48,38){$\circ$} 
\put(27,57){$\Box$} 
\put(48,58){$\circ$} 
\put(68,58){$\circ$} 
\put(8,78){$\star$} 
\put(27,77){$\Box$} 
\put(47,77){$\Box$} 
\put(68,78){$\star$}
\put(28,98){$\star$}
\put(50,60){\textcolor{red}{\vector(1,0){20}}}
\put(70,60){\textcolor{red}{\vector(0,1){20}}}
\put(140,30){\line(1,0){20}}
\put(120,50){\line(1,0){60}}
\put(100,70){\line(1,0){80}}
\put(100,90){\line(1,0){80}}
\put(120,110){\line(1,0){20}}
\put(100,70){\line(0,1){20}}
\put(120,50){\line(0,1){60}}
\put(140,30){\line(0,1){80}}
\put(160,30){\line(0,1){60}}
\put(180,50){\line(0,1){40}}
\put(147,37){$\Box$} 
\put(128,58){$\circ$} 
\put(147,57){$\Box$} 
\put(167,57){$\Box$} 
\put(108,78){$\star$} 
\put(128,78){$\circ$} 
\put(148,78){$\circ$} 
\put(168,78){$\star$}
\put(128,98){$\star$} 
\put(132,60){\textcolor{blue}{\vector(0,1){20}}}
\put(130,80){\textcolor{blue}{\vector(-1,0){20}}}
\put(220,30){\line(1,0){40}}
\put(220,50){\line(1,0){60}}
\put(200,70){\line(1,0){80}}
\put(200,90){\line(1,0){80}}
\put(260,110){\line(1,0){20}}
\put(200,70){\line(0,1){20}}
\put(220,30){\line(0,1){60}}
\put(240,30){\line(0,1){60}}
\put(260,30){\line(0,1){80}}
\put(280,50){\line(0,1){60}}
\put(228,38){$\bullet$} 
\put(248,38){$\circ$} 
\put(227,57){$\Box$} 
\put(248,58){$\circ$} 
\put(267,57){$\Box$} 
\put(208,78){$\bullet$}
\put(227,77){$\circ$}  
\put(247,77){$\Box$} 
\put(268,78){$\bullet$}
\put(268,98){$\bullet$}
\put(40,10){$O$}  
\put(140,10){$O'$}
\put(240,10){$O''$}    
\end{picture}  
\begin{picture}(290,130)
\put(-20,110){(C)} 
\put(20,30){\line(1,0){40}}
\put(20,50){\line(1,0){60}}
\put(0,70){\line(1,0){80}}
\put(0,90){\line(1,0){80}}
\put(20,110){\line(1,0){20}}
\put(0,70){\line(0,1){20}}
\put(20,30){\line(0,1){80}}
\put(40,30){\line(0,1){80}}
\put(60,30){\line(0,1){60}}
\put(80,50){\line(0,1){40}}
\put(28,38){$\star$} 
\put(48,38){$\circ$} 
\put(27,57){$\Box$} 
\put(48,58){$\circ$} 
\put(68,58){$\circ$} 
\put(8,78){$\star$} 
\put(27,77){$\Box$} 
\put(47,77){$\Box$} 
\put(68,78){$\star$}
\put(28,98){$\star$}
\put(50,60){\textcolor{red}{\vector(1,0){20}}}
\put(70,60){\textcolor{red}{\vector(0,1){20}}}
\put(70,80){\textcolor{red}{\vector(-1,0){20}}}
\put(140,30){\line(1,0){20}}
\put(120,50){\line(1,0){60}}
\put(100,70){\line(1,0){80}}
\put(100,90){\line(1,0){80}}
\put(120,110){\line(1,0){20}}
\put(100,70){\line(0,1){20}}
\put(120,50){\line(0,1){60}}
\put(140,30){\line(0,1){80}}
\put(160,30){\line(0,1){60}}
\put(180,50){\line(0,1){40}}
\put(147,37){$\Box$} 
\put(128,58){$\circ$} 
\put(147,57){$\Box$} 
\put(167,57){$\Box$} 
\put(108,78){$\star$} 
\put(128,78){$\circ$} 
\put(148,78){$\circ$} 
\put(168,78){$\star$}
\put(128,98){$\star$} 
\put(220,30){\line(1,0){40}}
\put(220,50){\line(1,0){60}}
\put(200,70){\line(1,0){80}}
\put(200,90){\line(1,0){80}}
\put(260,110){\line(1,0){20}}
\put(200,70){\line(0,1){20}}
\put(220,30){\line(0,1){60}}
\put(240,30){\line(0,1){60}}
\put(260,30){\line(0,1){80}}
\put(280,50){\line(0,1){60}}
\put(228,38){$\bullet$} 
\put(248,38){$\circ$} 
\put(227,57){$\Box$} 
\put(248,58){$\circ$} 
\put(267,57){$\Box$} 
\put(208,78){$\bullet$}
\put(227,77){$\circ$}   
\put(247,77){$\Box$} 
\put(268,78){$\bullet$}
\put(268,98){$\bullet$}
\put(40,10){$O$}  
\put(140,10){$O'$}
\put(240,10){$O''$}    
\end{picture}  
\end{center}
\label{eq.sect.object.1.ABC} 
\end{figure}
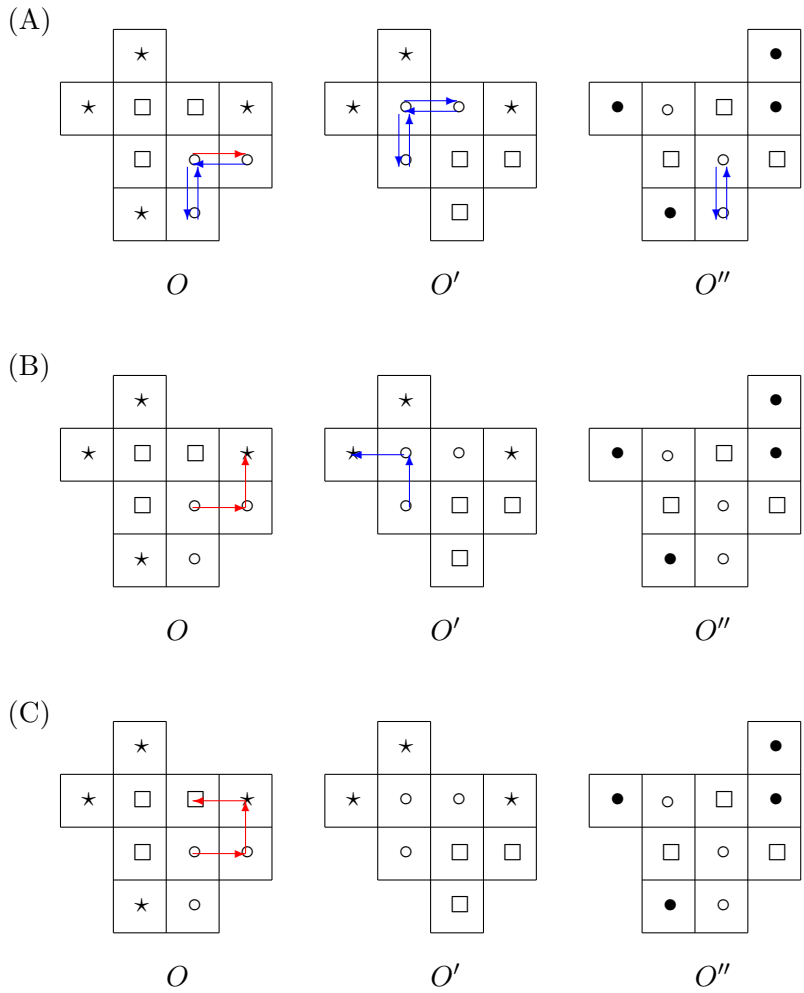
\newpage
\noindent
represented by the blue vectors on $O,$ $O',$ and $O''.$    
Therefore, the model can not identify the object that it 
has observed.  
In Figure \ref{eq.sect.object.1.ABC} (B), 
the red arrows represent the first and second 
real movement vectors. 
Then, these movements can not be distinguished from the movements 
represented by the blue vectors on $O'.$    
Therefore, although the model is aware that it did not observe $O'',$ 
it is unaware whether it has observed object $O$ or $O'.$   
In Figure \ref{eq.sect.object.1.ABC} (C), 
the red arrows represent the first, second, and third
real movement vectors. 
Then, the model is aware that it has observed $O,$ that is, 
it recognizes $O.$
When this is the case, 
it is said that the inference has 
converged onto a representation 
for the object $O.$ 
The convergence 
property has been investigated in detail in \cite{HAC} and \cite{LPAH}.

The inference by the Numenta model converges to an object, 
as the ambiguity regarding the object under observation decreases. 
The problem of inferring an object using observed data 
is an inverse problem. One of the well-known methods
for solving inverse problems is successive approximation 
such as gradient descent.
By contrast, the inference of the Numenta model can be 
considered as a Bayesian updating inference
(see \S \ref{sect.inference}).  

\section{Numenta model of the neocortex} 
\label{sect.model} 

\begin{figure}[H]
\caption{Numenta cortical column} 
\unitlength 1pt
\begin{center}
\begin{picture}(300,400)
\put(60,10){\line(1,0){220}}
\put(60,110){\line(1,0){220}}
\put(60,150){\line(1,0){220}}
\put(60,250){\line(1,0){220}}
\put(60,10){\line(0,1){100}}
\put(60,150){\line(0,1){100}}
\put(280,10){\line(0,1){100}}
\put(280,150){\line(0,1){100}}
\put(5,82){location} 
\put(5,70){layer} 
\put(5,250){sensory} 
\put(5,238){layer} 
\put(5,222){(input} 
\put(5,210){layer)} 
\put(5,330){output} 
\put(5,318){layer} 
\put(100,350){\textcolor{blue}{$\bullet$}}
\put(130,360){\textcolor{blue}{$\bullet$}}
\put(150,300){\textcolor{blue}{$\bullet$}}
\put(170,340){\textcolor{blue}{$\bullet$}}
\put(180,370){\textcolor{blue}{$\bullet$}}
\put(200,330){\textcolor{blue}{$\bullet$}}
\put(220,320){\textcolor{blue}{$\bullet$}}
\put(218,230){\textcolor{blue}{$\bullet$}}
\put(218,220){\textcolor{blue}{$\bullet$}}
\put(218,210){\textcolor{blue}{$\bullet$}}
\put(218,200){\textcolor{blue}{$\bullet$}}
\put(218,190){\textcolor{blue}{$\bullet$}}
\put(218,180){\textcolor{blue}{$\bullet$}}
\put(218,170){\textcolor{blue}{$\bullet$}}
\put(218,160){\textcolor{blue}{$\bullet$}}
\put(195,135){mini-column}
\put(0,50){\textcolor{blue}{\line(1,0){60}}}
\put(0,190){\textcolor{blue}{\line(1,0){60}}}
\put(100,150){\textcolor{blue}{\line(0,-1){40}}}
\put(160,150){\textcolor{blue}{\line(0,-1){40}}}
\put(60,50){\textcolor{blue}{\line(-1,1){10}}}
\put(60,50){\textcolor{blue}{\line(-1,-1){10}}}
\put(60,190){\textcolor{blue}{\line(-1,1){10}}}
\put(60,190){\textcolor{blue}{\line(-1,-1){10}}}
\put(100,150){\textcolor{blue}{\line(1,-1){10}}}
\put(100,150){\textcolor{blue}{\line(-1,-1){10}}}
\put(160,110){\textcolor{blue}{\line(-1,1){10}}}
\put(160,110){\textcolor{blue}{\line(1,1){10}}}
\put(20,40){$\bigcirc$\hspace*{-8pt}1}
\put(20,180){$\bigcirc$\hspace*{-8pt}3}
\put(86,122){$\bigcirc$\hspace*{-8pt}2}
\put(146,130){$\bigcirc$\hspace*{-8pt}4}
\put(90,160){\textcolor{red}{$D_{c,d}^{in}$}}
\put(150,95){\textcolor{red}{$D_{\gamma,d}^{loc}$}}
\put(68,188){\textcolor{red}{$W_t^{in}$}}
\put(100,300){\textcolor{red}{$f_{ijk}$}}
\put(160,235){\textcolor{red}{$f_{ijk}$}}
\put(88,368){\textcolor{red}{$D_{k,d}^{out}$}}
\put(230,368){\textcolor{red}{$D_{k,d}^{out}$}}
\put(140,30){\textcolor{blue}{$\bullet$}}
\put(150,30){\textcolor{blue}{$\bullet$}}
\put(160,30){\textcolor{blue}{$\bullet$}}
\put(170,30){\textcolor{blue}{$\bullet$}}
\put(145,40){\textcolor{blue}{$\bullet$}}
\put(155,40){\textcolor{blue}{$\bullet$}}
\put(165,40){\textcolor{blue}{$\bullet$}}
\put(175,40){\textcolor{blue}{$\bullet$}}
\put(150,50){\textcolor{blue}{$\bullet$}}
\put(160,50){\textcolor{blue}{$\bullet$}}
\put(170,50){\textcolor{blue}{$\bullet$}}
\put(180,50){\textcolor{blue}{$\bullet$}}
\put(155,60){\textcolor{blue}{$\bullet$}}
\put(165,60){\textcolor{blue}{$\bullet$}}
\put(175,60){\textcolor{blue}{$\bullet$}}
\put(185,60){\textcolor{blue}{$\bullet$}}
\put(200,50){module} 
\put(0,370){\line(1,0){80}}
\put(70,380){\textcolor{blue}{\line(1,-1){10}}}
\put(70,360){\textcolor{blue}{\line(1,1){10}}}
\put(30,375){$\bigcirc$\hspace*{-8pt}6}
\put(0,370){\textcolor{blue}{\line(1,1){10}}}
\put(0,370){\textcolor{blue}{\line(1,-1){10}}}
\put(260,370){\line(1,0){80}}
\put(330,380){\textcolor{blue}{\line(1,-1){10}}}
\put(330,360){\textcolor{blue}{\line(1,1){10}}}
\put(300,375){$\bigcirc$\hspace*{-8pt}6}
\put(260,370){\textcolor{blue}{\line(1,1){10}}}
\put(260,370){\textcolor{blue}{\line(1,-1){10}}}
\put(110,250){\line(0,1){40}}
\put(170,250){\line(0,1){40}}
\put(170,250){\textcolor{blue}{\line(-1,1){10}}}
\put(170,250){\textcolor{blue}{\line(1,1){10}}}
\put(110,290){\textcolor{blue}{\line(-1,-1){10}}}
\put(110,290){\textcolor{blue}{\line(1,-1){10}}}
\put(95,260){$\bigcirc$\hspace*{-8pt}5}
\put(155,270){$\bigcirc$\hspace*{-8pt}7}
\put(60,290){\line(1,0){220}}
\put(60,390){\line(1,0){220}}
\put(60,290){\line(0,1){100}}
\put(280,290){\line(0,1){100}}
\end{picture}
\end{center}    
\label{eq.subsect.hac_lpah.structure.1}
\end{figure}
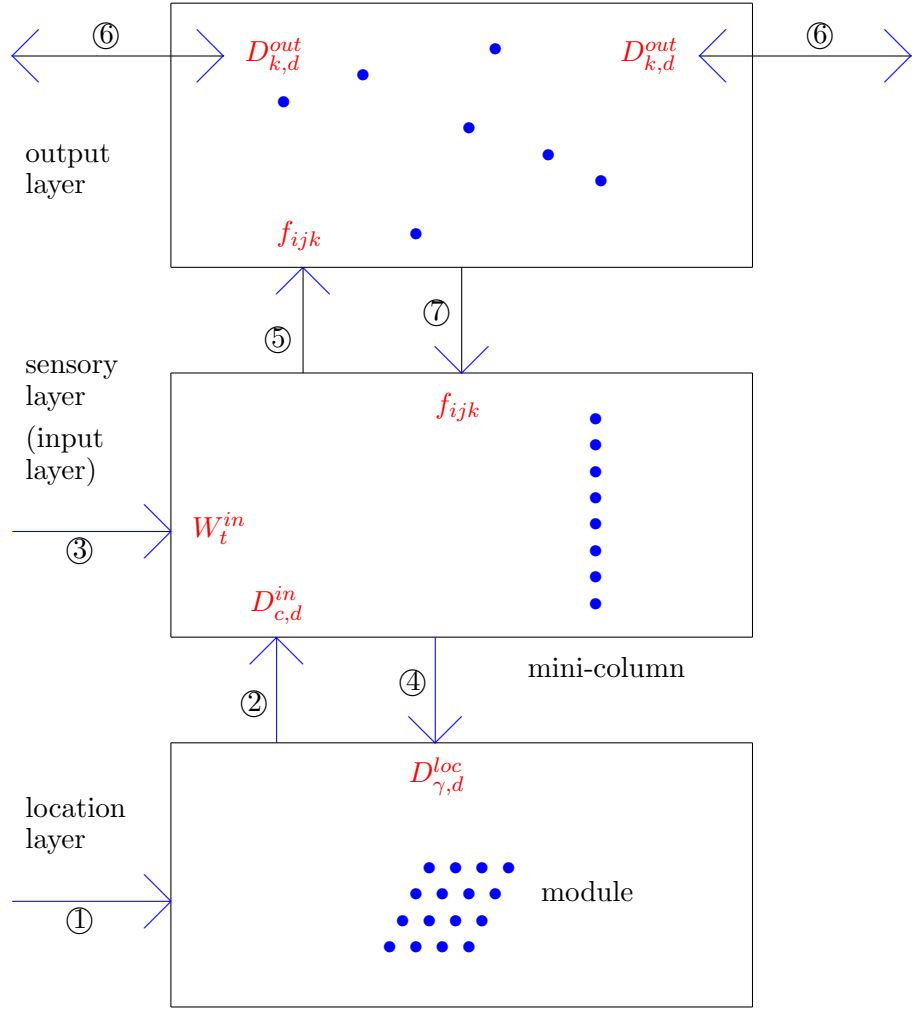

This section reviews the Numenta model 
of the structure of neocortex and learning and inference 
algorithms operating on the neocortex, 
based on \cite{HAC} and \cite{LPAH}. 
However, it is slightly changed, mainly for simplicity
(see Remarks \ref{remark.subsect.model.learning.1}
and \ref{remark.subsect.model.inference.2}).  
The neocortex comprises numerous cortical columns 
that are stacked vertically next to each other. 
All the cortical columns have the same structure and 
Figure \ref{eq.subsect.hac_lpah.structure.1} 
shows one of the cortical columns. 
Although each cortical column is said to comprise six horizontal layers, 
the cortical column of 
Figure \ref{eq.subsect.hac_lpah.structure.1} 
comprises the following three layers:
the output, sensory, and location layers 
corresponding to layers 2/3, 4, and 6a, respectively.
The sensory layer is also called the input layer
in \cite{HAC}. 
The model depicted in Figure 
\ref{eq.subsect.hac_lpah.structure.1} 
is obtained by combining the models in \cite{HAC} and \cite{LPAH}. 
The model in \cite{HAC} comprises only the output  
and sensory layers, and 
that in \cite{LPAH} comprises only the sensory  
and location layers. 
The model of Figure \ref{eq.subsect.hac_lpah.structure.1} 
also corresponds to the learning module in \cite{CLH} 
(see Figure 3 in \cite{CLH}). 

Each blue bullet in Figure 
\ref{eq.subsect.hac_lpah.structure.1} 
is an hierarchical temporal memory (HTM) neuron
(see \cite{HA} and \cite{HAD}).
HTM neurons are also called cells.
Each cell can be in one of the following 
three states: active, predictive, or inactive. 
The location layer comprises 
several grid cell modules, and 
each module comprises 
several cells arranged in a triangular lattice. 
In Figure \ref{eq.subsect.hac_lpah.structure.1},
only one module is depicted. 
The number of modules in 
the location layer of each cortical column is denoted by 
$N^{loc}$ and the number of cells in each module by   
$M^{loc}.$ 
In some simulations run in \cite{LPAH}, 
$N^{loc} = 10$ and $M^{loc} = 30 \times 30$ -- $40 \times 40.$ 
The sensory layer comprises  
several mini-columns, and 
each mini-column comprises  
several cells arranged in a line. 
In Figure \ref{eq.subsect.hac_lpah.structure.1},
only one mini-column is depicted. 
The number of mini-columns in 
the sensory layer of each cortical column is denoted by $N^{in}$ 
and the number of cells in each mini-column by $M^{in}.$  
In the simulations run in \cite{HAC} and 
\cite{LPAH}, 
$N^{in} = 150$ and $M^{in} = 16.$ 
The output layer has no internal structure such as modules or mini-columns.
In Figure \ref{eq.subsect.hac_lpah.structure.1},
only seven cells are depicted. 
The number of cells in 
the output layer of each cortical column is denoted by $N^{out}.$ 
In the simulations run in \cite{HAC}, $N^{out} = 4096.$ 

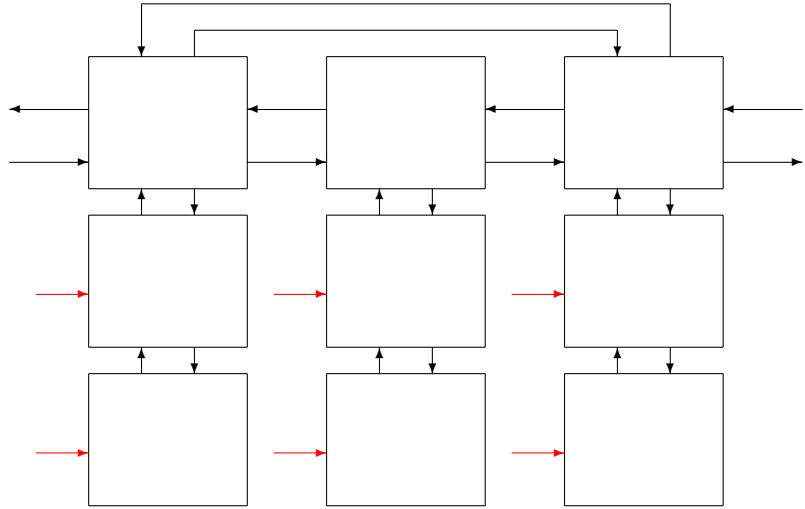
\begin{figure}[H]
\caption{Numenta cortical columns} 
\unitlength 1pt
\begin{center}
\begin{picture}(310,210)
\put(60,10){\line(1,0){60}} 
\put(150,10){\line(1,0){60}} 
\put(240,10){\line(1,0){60}} 
\put(60,60){\line(1,0){60}} 
\put(150,60){\line(1,0){60}} 
\put(240,60){\line(1,0){60}} 
\put(60,70){\line(1,0){60}} 
\put(150,70){\line(1,0){60}} 
\put(240,70){\line(1,0){60}} 
\put(60,120){\line(1,0){60}} 
\put(150,120){\line(1,0){60}} 
\put(240,120){\line(1,0){60}} 
\put(60,130){\line(1,0){60}} 
\put(150,130){\line(1,0){60}} 
\put(240,130){\line(1,0){60}} 
\put(60,180){\line(1,0){60}} 
\put(150,180){\line(1,0){60}} 
\put(240,180){\line(1,0){60}}
\put(60,10){\line(0,1){50}}  
\put(120,10){\line(0,1){50}}
\put(150,10){\line(0,1){50}}
\put(210,10){\line(0,1){50}}
\put(240,10){\line(0,1){50}}
\put(300,10){\line(0,1){50}}          
\put(60,70){\line(0,1){50}}  
\put(120,70){\line(0,1){50}}
\put(150,70){\line(0,1){50}}
\put(210,70){\line(0,1){50}}
\put(240,70){\line(0,1){50}}
\put(300,70){\line(0,1){50}}   
\put(60,130){\line(0,1){50}}  
\put(120,130){\line(0,1){50}}
\put(150,130){\line(0,1){50}}
\put(210,130){\line(0,1){50}}
\put(240,130){\line(0,1){50}}
\put(300,130){\line(0,1){50}}          
\put(40,30){\textcolor{red}{\vector(1,0){20}}}
\put(130,30){\textcolor{red}{\vector(1,0){20}}}
\put(220,30){\textcolor{red}{\vector(1,0){20}}}
\put(40,90){\textcolor{red}{\vector(1,0){20}}}
\put(130,90){\textcolor{red}{\vector(1,0){20}}}
\put(220,90){\textcolor{red}{\vector(1,0){20}}}
\put(80,60){\vector(0,1){10}}
\put(100,70){\vector(0,-1){10}}
\put(170,60){\vector(0,1){10}}
\put(190,70){\vector(0,-1){10}}
\put(260,60){\vector(0,1){10}}
\put(280,70){\vector(0,-1){10}}
\put(80,120){\vector(0,1){10}}
\put(100,130){\vector(0,-1){10}}
\put(170,120){\vector(0,1){10}}
\put(190,130){\vector(0,-1){10}}
\put(260,120){\vector(0,1){10}}
\put(280,130){\vector(0,-1){10}}
\put(30,140){\vector(1,0){30}}
\put(120,140){\vector(1,0){30}}
\put(210,140){\vector(1,0){30}}
\put(300,140){\vector(1,0){30}}
\put(60,160){\vector(-1,0){30}}
\put(150,160){\vector(-1,0){30}}
\put(240,160){\vector(-1,0){30}}
\put(330,160){\vector(-1,0){30}}
\put(100,190){\line(1,0){160}}
\put(100,180){\line(0,1){10}}
\put(260,190){\vector(0,-1){10}}
\put(80,200){\line(1,0){200}}
\put(80,200){\vector(0,-1){20}}
\put(280,180){\line(0,1){20}}
\end{picture}
\end{center}
\label{eq.subsect.hac_lpah.1} 
\end{figure}

The Numenta model is a discrete time model.  
The arrows in Figure \ref{eq.subsect.hac_lpah.structure.1}  
represents the flow of information between the cells.
In the inference, the one cycle of the ordered flow is 
$\bigcirc$\hspace*{-8pt}1 $\rightarrow$  
$\bigcirc$\hspace*{-8pt}2 $\rightarrow$  
$\bigcirc$\hspace*{-8pt}3 $\rightarrow$  
$\bigcirc$\hspace*{-8pt}5 $\rightarrow$    
$\bigcirc$\hspace*{-8pt}6 $\rightarrow$    
$\bigcirc$\hspace*{-8pt}7 $\rightarrow$
$\bigcirc$\hspace*{-8pt}4 $\rightarrow$        
$\bigcirc$\hspace*{-8pt}1\ . 
In the model proposed by 
\cite{LPAH}, it is  
$\bigcirc$\hspace*{-8pt}1 $\rightarrow$  
$\bigcirc$\hspace*{-8pt}2 $\rightarrow$  
$\bigcirc$\hspace*{-8pt}3 $\rightarrow$  
$\bigcirc$\hspace*{-8pt}4 $\rightarrow$    
$\bigcirc$\hspace*{-8pt}1\ , and 
in the model proposed by  \cite{HAC}, it is 
$\bigcirc$\hspace*{-8pt}2 $\rightarrow$  
$\bigcirc$\hspace*{-8pt}3 $\rightarrow$  
$\bigcirc$\hspace*{-8pt}5 $\rightarrow$  
$\bigcirc$\hspace*{-8pt}6 $\rightarrow$  
$\bigcirc$\hspace*{-8pt}7 $\rightarrow$  
$\bigcirc$\hspace*{-8pt}2\ . 
These two flows are sub-flows of the flow shown in Figure 
\ref{eq.subsect.hac_lpah.structure.1}. 
The information flows in learning are similar to the above. 
Steps $\bigcirc$\hspace*{-8pt}1\ to $\bigcirc$\hspace*{-8pt}4 
\ correspond to the stages 1 to 4 in \cite{LPAH}.   
Arrow $\bigcirc$\hspace*{-8pt}1\ is called the motor input
in \cite{LPAH},  
and $\bigcirc$\hspace*{-8pt}3\ is called the sensory input
in \cite{HAC} and \cite{LPAH}. 
These inputs originate from outside the cortical column, and 
the motor input is either conscious or unconscious.
Arrow $\bigcirc$\hspace*{-8pt}6\ shows 
the internal flow of the output layer of a cortical column and
the flow between the output layers of cortical columns. 
Figure \ref{eq.subsect.hac_lpah.1} illustrates 
three cortical columns.  
Different cortical columns may receive the same type of sensory inputs,
and they may also receive different types of sensory inputs,  
such as shape and color in vision. 
Therefore, the Numenta model can handle multimodal sensory inputs. 

Let $v$ be a vector or tensor. If each component of $v$ is either 0 or 1,
we refer to $v$ as a binary vector or tensor. 
For a binary vector or tensor $v,$ 
the number of 1s in the components of $v$ 
is denoted by $\natural v.$
The inner product of vectors $u$ and $v$ of the same dimension is 
denoted by $u\iprod v.$ 
Let $N^c$ be the number of the considered cortical columns.
We assume that the values of 
$N^{loc},$ $M^{loc},$ $N^{in},$ $M^{in},$ and $N^{out}$  
are equal for all considered cortical columns. 
Let $D_{c,d}^{in},$
$D_{\gamma,d}^{loc},$
and $D_{k,d}^{out}$
be binary vectors, and 
$F = (f_{ijk})$ a binary tensor, where 
\[
\dim D_{c,d}^{in} = N^{loc} M^{loc}, 
\hspace{10pt} 
\dim D_{\gamma,d}^{loc} = N^{in} M^{in},
\hspace{10pt} 
\dim D_{k,d}^{out} = N^c N^{out},
\hspace{10pt} 
\dim F = N^{in} M^{in} N^{out}.  
\]
Vector $D_{c,d}^{in}$ represents 
a dendritic segment $d$ of 
a cell $c$ in the sensory layer, 
$D_{\gamma,d}^{loc}$ represents 
a dendritic segment $d$ of 
a cell $\gamma$ in the location layer, 
$D_{k,d}^{out}$ represents 
a dendritic segment $d$ of 
a cell $k$ in the output layer,   
and $f_{ijk}$ represents 
the pair of a cell $j$ in mini-column $i$ of the sensory layer
and a cell $k$ in the output layer. 
The components of $D_{c,d}^{in}$ correspond to all the cells in  
the location layer
of the cortical column containing $c,$
those of $D_{\gamma,d}^{loc}$ 
correspond to all the cells in the sensory layer
of the cortical column containing $\gamma,$ and
those of $F$ correspond to all the pairs of 
the cells in the sensory  
and output layers in the same cortical column.
The components of $D_{k,d}^{out}$ correspond to all the cells in  
the output layers of all considered cortical columns. 
Each component of 
$D_{c,d}^{in},$ $D_{\gamma,d}^{loc},$
$D_{k,d}^{out},$ and $F$ represents the connections between specified cells;
for instance, if and only if a component of $D_{c,d}^{in}$ is 1, 
a connection exists between the cell in the location layer 
represented by this component and the segment $d$ of cell $c.$ 
All the capabilities of the Numenta model are realized by these connections.

In \S \ref{subsect.model.learning} and 
\S \ref{subsect.model.inference},  
we consider learning and inference/recognition algorithms 
for objects. 
Figure \ref{eq.subsect.parameters.example.1}  
shows an example of such an object.
This object $O$ comprises ten pairs of (location, feature).
When a cortical column observes or recalls $O,$
the location is specified by active cells in the location layer, and  
the feature is specified by active cells in the sensory layer.
Each module in the location layer 
acts as a reference frame (or coordinate frame) 
of the locations on the object under consideration.  
This is emphasized in \cite{H2} and \cite{LPAH}. 
According to \cite{H2} and \cite{LPAH}, 
the information flow in the model proposed by \cite{LPAH} 
is fundamentally sufficient for  
learning, inferring, and recognizing any (simple) object. 
If an object is complex to be recognized by only one cortical column, 
the connections $\bigcirc$\hspace*{-8pt}6\ 
between the output layers of the cortical columns assist 
in recognizing this object. 
According to \cite{CLH}, 
each learning module can recognize objects, 
and multiple learning modules can recognize more complex objects 
at a faster rate 
through voting and a hierarchical structure.

Algorithm \ref{algorithm.subsect.model.learning.1} is a 
learning algorithm and 
Algorithm \ref{algorithm.subsect.model.inference.1} is an  
inference algorithm, based on \cite{HAC} and \cite{LPAH}. 
In the author's opinion,
some steps omitted in the algorithms of \cite{HAC} and \cite{LPAH} 
are added to Algorithms 
\ref{algorithm.subsect.model.learning.1} 
and \ref{algorithm.subsect.model.inference.1}, 
and some steps are changed, mainly for simplicity.
In Algorithms 
\ref{algorithm.subsect.model.learning.1} 
and \ref{algorithm.subsect.model.inference.1},  
$A_t^{loc}$ and $A_t^{in}$ are binary vectors 
such that 
\[
\dim A_t^{loc} = N^{loc}M^{loc}, \hspace{10pt} 
\dim A_t^{in} = N^{in} M^{in}, 
\]
and the components of $A_t^{loc}$ and $A_t^{in}$    
correspond to the cells in the location  
and sensory layers of the considered cortical column 
at time $t,$ respectively.    
If and only if a component of $A_t^{loc}$ or $A_t^{in}$  
is 1, the corresponding cell is active.
In the following, 
$A_t^{loc}$ and $A_t^{in}$ are identified with
the sets of all cells in 
the location and sensory layers, respectively.
When a location on an object is observed, the feature $f$ at this location 
provides a sensory input to the sensory layer, some mini-columns 
in the sensory layer are selected,  
and some cells in these mini-columns are activated. 
The set of such selected mini-columns 
is denoted by $W^{in}(f).$ 
Note that $W^{in}(f)$ is sparse, 
that is, $\sharp W^{in}(f)$ 
is significantly lower than $N^{in},$
where $\sharp S$ for a set $S$ 
is the number of elements of $S.$ 
In \cite{HAD}, 
$W^{in}(f)$ is called the sparse distributed representation (SDR) of $f.$  
In Algorithms 
\ref{algorithm.subsect.model.learning.1} 
and \ref{algorithm.subsect.model.inference.1}, 
$W^{in}(f)$ at time $t$ is denoted by $W^{in}_t = W^{in}_t(f).$ 

\subsection{Learning} 
\label{subsect.model.learning} 

Algorithm \ref{algorithm.subsect.model.learning.1} is 
a learning algorithm obtained by combining such algorithms  
of \cite{HAC} and \cite{LPAH}. 
Algorithm \ref{algorithm.subsect.model.learning.1} learns an 
object $O$ by observing and sensing 
pairs (location, feature) on $O$ individually. 
\begin{algorithm} 
\label{algorithm.subsect.model.learning.1} 
{\rm 
(Numenta learning algorithm)

This algorithm runs on each cortical column. 
In this algorithm, steps 
\ref{step.algorithm.subsect.model.learning.1.new6} to 
\ref{step.algorithm.subsect.model.learning.1.12}
are repeated from the second round onwards.
If $\pi_{c,t}^{in} = 0$ for every $c$ in 
(\ref{eq.algorithm.subsect.model.learning.1.5}), 
steps 
\ref{step.algorithm.subsect.model.learning.1.11} and   
\ref{step.algorithm.subsect.model.learning.1.12}  
are the same as step 
\ref{step.algorithm.subsect.model.learning.1.5}.   
The positive constant $\theta_b^{in}$ in 
(\ref{eq.algorithm.subsect.model.learning.1.5})  
is a threshold.  
The symbol ``$|$'' in 
(\ref{eq.algorithm.subsect.model.learning.1.2}), 
(\ref{eq.algorithm.subsect.model.learning.1.3}), 
and (\ref{eq.algorithm.subsect.model.learning.1.1}) 
is designated as bitwise OR. 
\begin{enumerate} 
\item
\label{step.algorithm.subsect.model.learning.1.1} 
Set $D_{c,d}^{in} = 0$ for every $(c,d),$
$D_{\gamma,d}^{loc} = 0$ for every $(\gamma,d),$
$D_{k,d}^{out} = 0$ for every $(k,d),$  
and $f_{ijk} = 0$ for every $(i,j,k).$  
\item
\label{step.algorithm.subsect.model.learning.1.2} 
For the object $O,$ 
select a binary vector $A_O^{out}$ of dimension $N^{out}$ at random,
that is, the values of the components of $A_O^{out}$ are 
determined at random.   
However, $A_O^{out}$ must be sparse, that is, 
$\natural A_O^{out}$ 
must be much less than $\dim A_O^{out}.$ 
This $A_O^{out}$ 
is fixed throughout 
this algorithm.
Denote by $\overline{A}_O^{out}$ the $N^cN^{out}$ dimensional vector 
obtained by concatenating $A_O^{out}$s of   
all considered cortical columns.

The components of $A_O^{out}$ 
correspond to all cells of the output layer. 
If and only if a component of $A_O^{out}$  
is 1, the corresponding cell is active.
In the following, 
$A_O^{out}$ is identified with 
the set of all cells in the output layer.
\item
\label{step.algorithm.subsect.model.learning.1.3} 
For every active cell $k \in A_O^{out},$ 
select a dendritic segment $d$ 
of $k$ at random and 
set $D_{k,d}^{out} = \overline{A}_O^{out}.$ 
Vector $D_{k,d}^{out}$ 
is fixed throughout this algorithm. 
\item
\label{step.algorithm.subsect.model.learning.1.4} 
Set $t = 0$ and start observing $O.$ 
From each module $i$ in the location layer, 
randomly select one cell 
and make it active.
Thus, the initial value of the vector of 
$A_t^{loc}$ is set.
  
The active cell in module $i$ at time $t$ 
represents a position vector $\vec{\phi}_{i,t}$ 
in the reference frame given by module $i.$ 
The set of vectors $\Phi_t := \left\{\vec{\phi}_{i,t}\right\}$  
corresponds to the current observation location on $O.$ 
\item 
\label{step.algorithm.subsect.model.learning.1.5} 
This step is stage $\bigcirc$\hspace*{-8pt}3\  
in Figure \ref{eq.subsect.hac_lpah.structure.1}. 
Sense the feature of $O$ at the location in step  
\ref{step.algorithm.subsect.model.learning.1.4}. 
For the sensory input from the feature, select a set of mini-columns 
$W^{in}_t$ of the sensory layer as follows.
If the input has been observed in a previous learning, 
let $W_t^{in}$ be the mini-columns 
selected then. If not, 
randomly select $W_t^{in}$ such that $\sharp W_t^{in} \ll N^{in}.$ 
Select one cell from each mini-column of $W_t^{in}$ at random 
and make this cell active. 
Thus, the initial value of the vector of 
$A_t^{in}$ is set. 
\item
\label{step.algorithm.subsect.model.learning.1.new6}
This step is stage $\bigcirc$\hspace*{-8pt}2\ .   
For every active cell $c$ in the sensory layer, 
select a dendritic segment $d$ of $c$ at random.   
It is fixed throughout this algorithm.
For every such pair $(c,d),$ 
update $D_{c,d}^{in}$ by  
\begin{equation}
D_{c,d}^{in} 
:= D_{c,d}^{in}\ \left| A_t^{loc} \right..
\label{eq.algorithm.subsect.model.learning.1.2} 
\end{equation}
This is equation (9) of \cite{LPAH}. 
\item
\label{step.algorithm.subsect.model.learning.1.8}
This step is stages $\bigcirc$\hspace*{-8pt}5  
\ and $\bigcirc$\hspace*{-8pt}7\ . 
For every active cell $k \in A_O^{out},$ 
randomly select some active cells $\{c_{ij}\}$ 
in the sensory layer such that 
$\sharp \{c_{ij}\} < \sharp W^{in}_t.$ 
Set $\gamma_{ijk} = 1$ and update $f_{ijk}$ by 
\begin{equation}
f_{ijk} := f_{ijk}\ \left| \gamma_{ijk} \right..
\label{eq.algorithm.subsect.model.learning.1.3} 
\end{equation}  
\item
\label{step.algorithm.subsect.model.learning.1.6}
This step is stage $\bigcirc$\hspace*{-8pt}4\ .  
For every active cell $\gamma$ in the location layer, 
select a dendritic segment $d$ of $\gamma$ at random.  
It is fixed throughout this algorithm.
For every such pair $(\gamma,d),$ 
update $D_{\gamma,d}^{loc}$ by  
\begin{equation}
D_{\gamma,d}^{loc}
:= D_{\gamma,d}^{loc}\ \left| A_t^{in} \right..
\label{eq.algorithm.subsect.model.learning.1.1} 
\end{equation}
This is equation (8) of \cite{LPAH}. 
\item
If the observation of $O$ is finished, stop this algorithm.
Otherwise, set $t := t + 1$ and go to the next step.   
\item
\label{step.algorithm.subsect.model.learning.1.10} 
This step is stage $\bigcirc$\hspace*{-8pt}1\ .
Change the observation location on $O$ by motor input. 
This motor input is represented  
by a vector $\vec{\delta}_{i,t}$ 
in each module $i$ 
of the location layer, and we obtain 
\[
\Phi_t = \left\{\vec{\phi}_{i,t} := \vec{\phi}_{i,t-1} 
+ \vec{\delta}_{i,t}\right\}, 
\]
where the addition 
$\vec{\phi}_{i,t-1} + \vec{\delta}_{i,t}$ is considered on the torus made 
from the lattice of module $i.$ 
Make all cells in $\Phi_t$ active and the other cells inactive.  
Thus, $A_t^{loc}$ is updated. 

The active cell in module $i$ 
represents a position vector $\vec{\phi}_{i,t}.$ 
The set $\Phi_t$  
corresponds to the current observation location on $O.$ 

Not only a vector that represents a movement on an object,
such as the red arrow in Figure \ref{eq.subsect.parameters.example.1},
but also a vector in the location layer that represents a motor input,
such as $\vec{\delta}_{i,t},$ is also called a movement vector.
\item
\label{step.algorithm.subsect.model.learning.1.11}
This step is stage $\bigcirc$\hspace*{-8pt}2\ . 
For every cell $c$ in the sensory layer, 
calculate
\begin{equation}
\pi_{c,t}^{in} :=  
\left\{
\begin{array}{ccl} 
1 & &\exists d:\ D_{c,d}^{in} 
\iprod A_t^{loc} \geq \theta_b^{in} 
\\
[5pt] 
0 & & \mbox{otherwise}. 
\end{array} 
\right. 
\label{eq.algorithm.subsect.model.learning.1.5} 
\end{equation}
If and only if $\pi_{c,t}^{in} = 1,$ the cell $c$ is predictive. 
If the current location is a location that has not been visited before, 
then $\pi_{c,t}^{in} = 0$ for almost all cells $c.$    
\item
\label{step.algorithm.subsect.model.learning.1.12}
This step is stage $\bigcirc$\hspace*{-8pt}3\ .  
Sense the feature of $O$ at the location in step  
\ref{step.algorithm.subsect.model.learning.1.10},
get sensory input, and select $W^{in}_t$ as in step 
\ref{step.algorithm.subsect.model.learning.1.5}.  
For every cell $c = (ij)$ in the sensory layer
(the $j$-th cell in the $i$-th mini-column), 
calculate the activity of $c$: 
\[
a_{ij,t}^{in} := 
\left\{
\begin{array}{lll} 
1 & & \mbox{if}\ i \in W_t^{in}\ \mbox{and}\ \pi_{ij,t}^{in} = 1 
\\
[5pt]
\ast & & \mbox{if}\ i \in W_t^{in}\ \mbox{and}\
\forall k \in \mbox{mini-column $i$}, 
\pi_{ik,t}^{in} = 0
\\
[5pt] 
0 & & \mbox{otherwise}, 
\end{array}
\right.
\]
where $\ast = 1$ for only one cell $j$ that is randomly selected 
from the $i$-th mini-column 
and $\ast = 0$ for the other every cell $j.$ 
If and only if $a_{ij,t}^{in} = 1,$ the cell $c = (ij)$ is active.   
Thus, $A_t^{in}$ is updated. 
Then, go back to step 
\ref{step.algorithm.subsect.model.learning.1.new6}. 
\end{enumerate} 
}
\end{algorithm} 

\begin{remark} 
\label{remark.subsect.model.learning.1} 
\hspace{5pt} 
{\rm
Compared with the learning algorithms in \cite{HAC} and 
\cite{LPAH}, 
Algorithm \ref{algorithm.subsect.model.learning.1} 
is simplified as follows:  
\begin{itemize}
\item
In the learning algorithm of \cite{HAC}, 
the synaptic permanence values are used  
for $D_{c,d}^{in},$ $D_{k,d}^{out},$ and $f_{ijk}$  
based on Hebbian-style adaptation
(see (6), (7), and (8) in \cite{HAC}). 
In contrast, 
in the learning algorithm of \cite{LPAH},
they are not used for $D_{c,d}^{in}$ and $D_{\gamma,d}^{loc}$  
as shown in 
(\ref{eq.algorithm.subsect.model.learning.1.2}) and   
(\ref{eq.algorithm.subsect.model.learning.1.1}), respectively. 
For simplicity,  
Algorithm \ref{algorithm.subsect.model.learning.1} 
does not use synaptic permanence values
for $D_{c,d}^{in},$ $D_{k,d}^{out},$ and $f_{ijk}$ either.
In particular, learning $D_{k,d}^{out}$ is performed only once, at step 
\ref{step.algorithm.subsect.model.learning.1.3}.
Note that, on page 5 in \cite{HAC}, the following is stated: 
``\textit{The output layer learns representations corresponding to
objects. When the network first encounters a new object, a sparse
set of cells in the output layer is selected to represent the new
object. These cells remain active while the system senses the
object at different locations.}''
Based on this, we maintain $A_O^{out}$ fixed throughout learning. 
\item
In \cite{LPAH}, 
the activity in the location layer is considered for 
not a cell but a bump of cells, and 
the structure of the reference frame and the lengths and angles 
of movement vectors are precisely defined.  
In the present study, the activity is considered 
only for a cell 
and movement vector $\vec{\delta}_{i,t}$ is used  
for simplicity. 
The important ideas  
of the modules acting as reference frames are   
explained in 
\cite{H2}, \cite{HLKPA}, \cite{LLA}, and \cite{LPAH}.  
\end{itemize}
}
\end{remark} 

\begin{remark} 
\label{remark.subsect.model.learning.3} 
\hspace{5pt} 
{\rm
We make some remarks regarding  
Algorithm \ref{algorithm.subsect.model.learning.1}.  
\begin{itemize}
\item
The Numenta model can handle multimodal information 
through connections between cortical columns 
via $\{D_{k,d}^{out}\}.$ 
Connections $\{D_{k,d}^{out}\}$ and $\{f_{ijk}\}$ 
realize associative memory. 
\item
For an object $O,$ vector $A_O^{out}$ 
is an SDR of $O.$ 
Therefore, if $O$ and $O'$ are different objects, 
$A_O^{out} \iprod A_{O'}^{out}$ is expected to be 
approximately zero. 
Additionally, learning a new object  
is expected not to result in catastrophic forgetting. 
\item
In \cite{HAC}, 
$\natural A_O^{out}$ 
in step \ref{step.algorithm.subsect.model.learning.1.2} 
typically satisfies $40 \leq \natural A_O^{out}  
\ll \dim A_O^{out} = 4096.$ 
In \cite{HAC} and \cite{HACS}, 
$\sharp W^{in}_t$ 
in steps \ref{step.algorithm.subsect.model.learning.1.5} 
and \ref{step.algorithm.subsect.model.learning.1.12} 
and
$\sharp\{c_{ij}\}$ 
in step \ref{step.algorithm.subsect.model.learning.1.8} 
are constants throughout learning, 
and their typical values are 
$10 = \sharp W^{in}_t \ll N^{in} = 150$  
and $\sharp\{c_{ij}\} = 5$ -- $8.$   
In the present study, the values of $\natural A_O^{out},$ $\sharp W^{in}_t,$ 
and $\sharp\{c_{ij}\}$ are assumed to be equal for all objects, 
features, and times.
\item
In Algorithm \ref{algorithm.subsect.model.learning.1}, 
overlaps of the learned cells corresponding 
to different objects probably exist 
because of random selections.  
In real learning, 
some noises that interfere with it probably exist.
See \cite{HAC}, \cite{HACS}, and \cite{LPAH}  
for the capacity for representing locations and features, and 
noise robustness.
\end{itemize}
}
\end{remark} 

\begin{figure}[H]
\caption{Learned connections between cells} 
\unitlength 1pt
\begin{center}
\begin{picture}(360,440)
\put(60,30){\line(1,0){120}}
\put(220,30){\line(1,0){120}}
\put(60,130){\line(1,0){120}}
\put(220,130){\line(1,0){120}}
\put(60,170){\line(1,0){120}}
\put(220,170){\line(1,0){120}}
\put(60,270){\line(1,0){120}}
\put(220,270){\line(1,0){120}}
\put(60,310){\line(1,0){120}}
\put(220,310){\line(1,0){120}}
\put(60,410){\line(1,0){120}}
\put(220,410){\line(1,0){120}}
\put(60,30){\line(0,1){100}}
\put(90,30){\line(0,1){100}}
\put(120,30){\line(0,1){100}}
\put(150,30){\line(0,1){100}}
\put(180,30){\line(0,1){100}}
\put(60,170){\line(0,1){100}}
\put(84,170){\line(0,1){100}}
\put(108,170){\line(0,1){100}}
\put(132,170){\line(0,1){100}}
\put(156,170){\line(0,1){100}}
\put(180,170){\line(0,1){100}}
\put(60,310){\line(0,1){100}}
\put(180,310){\line(0,1){100}} 
\put(220,30){\line(0,1){100}}
\put(250,30){\line(0,1){100}}
\put(280,30){\line(0,1){100}}
\put(310,30){\line(0,1){100}}
\put(340,30){\line(0,1){100}}
\put(220,170){\line(0,1){100}}
\put(244,170){\line(0,1){100}}
\put(268,170){\line(0,1){100}}
\put(292,170){\line(0,1){100}}
\put(316,170){\line(0,1){100}}
\put(340,170){\line(0,1){100}}
\put(220,310){\line(0,1){100}} 
\put(340,310){\line(0,1){100}} 
\put(5,82){location} 
\put(5,70){layer} 
\put(5,219){\textcolor{red}{$W_t^{in}$} in} 
\put(5,207){sensory} 
\put(5,195){layer} 
\put(5,364){\textcolor{red}{$A_O^{out}$} in} 
\put(5,352){output} 
\put(5,340){layer} 
\put(73,80){$\textcolor{red}{\bullet}$} 
\put(103,80){$\textcolor{red}{\bullet}$} 
\put(133,80){$\textcolor{red}{\bullet}$} 
\put(163,80){$\textcolor{red}{\bullet}$} 
\put(233,80){$\textcolor{red}{\bullet}$} 
\put(263,80){$\textcolor{red}{\bullet}$} 
\put(293,80){$\textcolor{red}{\bullet}$} 
\put(323,80){$\textcolor{red}{\bullet}$} 
\put(70,220){$\textcolor{red}{\bullet}$} 
\put(93,220){$\textcolor{red}{\bullet}$} 
\put(118,220){$\textcolor{red}{\bullet}$} 
\put(141,220){$\textcolor{red}{\bullet}$}
\put(165,220){$\textcolor{red}{\bullet}$}  
\put(230,220){$\textcolor{red}{\bullet}$} 
\put(253,220){$\textcolor{red}{\bullet}$} 
\put(278,220){$\textcolor{red}{\bullet}$} 
\put(301,220){$\textcolor{red}{\bullet}$}
\put(325,220){$\textcolor{red}{\bullet}$}  
\put(118,340){$\textcolor{red}{\bullet}$} 
\put(253,340){$\textcolor{red}{\bullet}$} 
\put(278,340){$\textcolor{red}{\bullet}$} 
\put(301,340){$\textcolor{red}{\bullet}$} 
\put(253,382){$\textcolor{red}{\bullet}$} 
\put(278,382){$\textcolor{red}{\bullet}$} 
\put(301,382){$\textcolor{red}{\bullet}$} 
\put(301,382){$\textcolor{red}{\bullet}$} 
\put(352,382){$\textcolor{red}{\bullet}$} 
\put(280,344){\textcolor{blue}{\line(0,1){38}}}
\put(280,344){\textcolor{blue}{\line(-2,3){24}}}
\put(280,344){\textcolor{blue}{\line(2,3){24}}}
\put(280,344){\textcolor{blue}{\line(2,1){72}}}
\put(278,344){\textcolor{blue}{\line(-1,0){20}}}
\put(282,344){\textcolor{blue}{\line(1,0){20}}}
\put(235,360){\textcolor{blue}{$\cdots\cdots$}}
\put(70,140){totally connected}
\put(230,150){totally connected}
\put(230,420){totally connected}
\put(120,218){\textcolor{blue}{\line(-1,-2){30}}}
\put(120,218){\textcolor{blue}{\line(-1,-3){20}}}
\put(120,218){\textcolor{blue}{\line(1,-2){30}}}
\put(120,218){\textcolor{blue}{\line(1,-3){20}}}
\put(265,82){\textcolor{blue}{\line(-1,2){30}}}
\put(265,82){\textcolor{blue}{\line(0,1){60}}}
\put(265,82){\textcolor{blue}{\line(1,3){20}}}
\put(265,82){\textcolor{blue}{\line(1,2){30}}}
\put(265,82){\textcolor{blue}{\line(2,3){40}}}
\put(120,338){\textcolor{blue}{\line(0,-1){114}}}
\put(117,338){\textcolor{blue}{\line(-1,-3){38}}}
\put(123,338){\textcolor{blue}{\line(1,-3){38}}}
\put(5,290){randomly selected}
\put(280,224){\textcolor{blue}{\line(0,1){72}}}
\put(280,224){\textcolor{blue}{\line(-1,2){36}}}
\put(280,224){\textcolor{blue}{\line(-1,3){24}}}
\put(280,224){\textcolor{blue}{\line(1,2){36}}}
\put(280,224){\textcolor{blue}{\line(1,3){24}}}
\put(170,145){\textcolor{red}{$\Uparrow$}}
\put(170,285){\textcolor{red}{$\Uparrow$}}
\put(330,145){\textcolor{red}{$\Downarrow$}}
\put(330,285){\textcolor{red}{$\Downarrow$}}
\put(190,340){\textcolor{red}{$\Longrightarrow$}}
\put(190,80){\textcolor{red}{$\Longleftarrow$}}
\put(165,35){$\fbox{1}$}
\put(165,175){$\fbox{2}$}
\put(165,315){$\fbox{3}$}
\put(325,35){$\fbox{6}$}
\put(325,175){$\fbox{5}$}
\put(325,315){$\fbox{4}$}
\end{picture}
\end{center}
\label{eq.subsect.hac_lpah.relation.connection.1}
\end{figure}
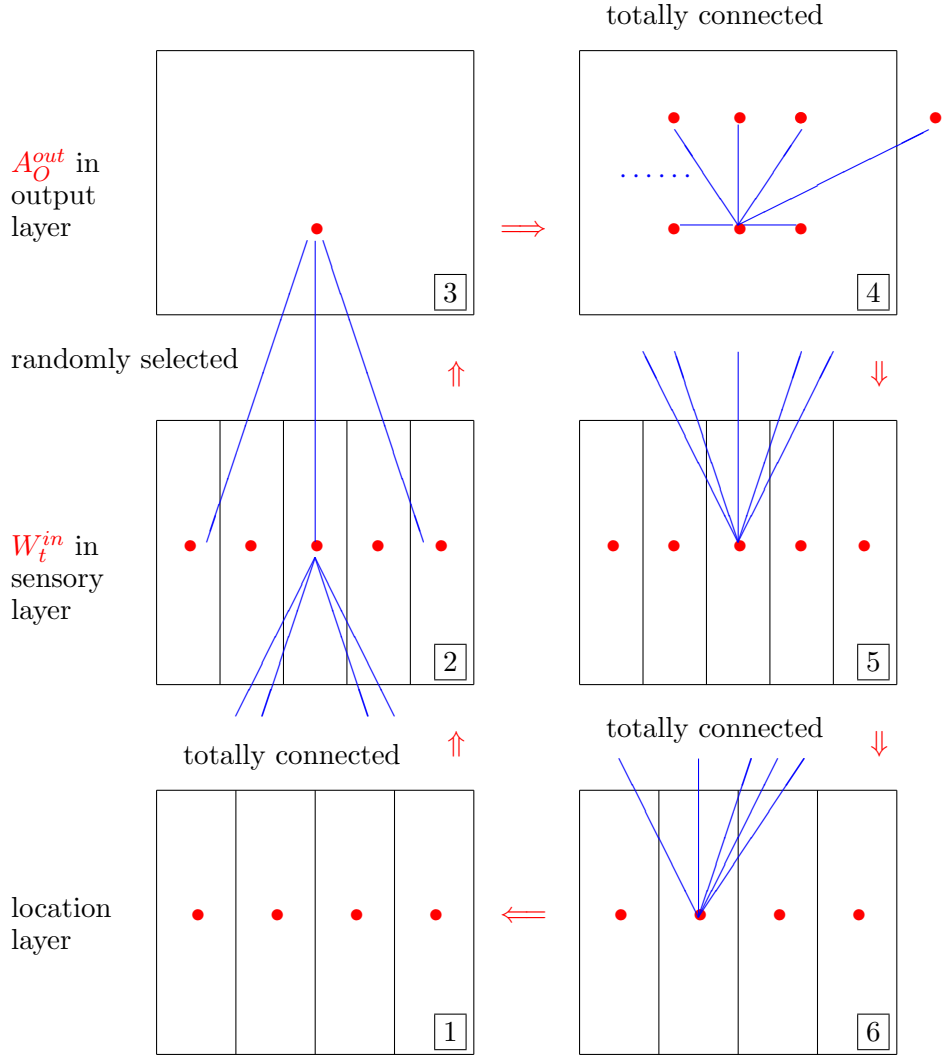

An example of 
the connections between cells obtained using the learning algorithm 
\ref{algorithm.subsect.model.learning.1} 
is illustrated in Figure 
\ref{eq.subsect.hac_lpah.relation.connection.1}. 
The figures of two cortical columns in Figure 
\ref{eq.subsect.hac_lpah.relation.connection.1} 
represent the same cortical column: 
on the left, information flows upwards, 
whereas on the right, information flows downwards.  
In $\fbox{1}$ and $\fbox{6},$ the four rectangles 
represent the modules, and the four red points represent 
the cells corresponding to a location. 
The five rectangles in $\fbox{2}$ and $\fbox{5}$ 
represent the mini-columns in $W^{in}_t$ for the feature sensed 
at the location of $\fbox{1}$ and $\fbox{6}.$
The five red points represent the cells selected 
from each mini-column in $W^{in}_t.$ 
For each cell in $A_O^{out},$ three cells were randomly selected  
from five active cells in $W^{in}_t.$  
Algorithm \ref{algorithm.subsect.model.learning.1} 
creates total connections between all the cells
representing the current location and all the cells representing 
the corresponding feature.
This algorithm also creates connections between
all the cells in the output layer that represent the same object.
However, these connections are not required 
to be total connections and
probabilistic connections are also possible. 

\subsection{Inference, prediction and recognition} 
\label{subsect.model.inference} 

Algorithm \ref{algorithm.subsect.model.inference.1} is 
an inference algorithm
for objects learned by Algorithm \ref{algorithm.subsect.model.learning.1}.  
It is obtained by combining such algorithms  
of \cite{HAC} and \cite{LPAH}. 
Algorithm \ref{algorithm.subsect.model.inference.1} makes 
inference by observing and sensing 
pairs (location, feature) on an object $O$ individually, 
in the same manner as in Algorithm \ref{algorithm.subsect.model.learning.1}.  
(In \S 2.2 of \cite{CLH}, it is stated that 
``\textit{there is no clear distinction between learning and
inference}.'')
Let $A_t^{out}$ be an $N^{out}$-dimensional binary vector 
such that the components of $A_t^{out}$ 
correspond to the cells of the output layer at time $t.$
If and only if a component of $A_t^{out}$ 
is 1, the corresponding cell is active.
In the following, $A_t^{out}$ is identified as 
the set of all cells in 
the output layer.
The $N^cN^{out}$-dimensional vector 
obtained by concatenating $A_t^{out}$s of   
all considered cortical columns
is denoted by $\overline{A}_t^{out}.$ 

\begin{algorithm} 
\label{algorithm.subsect.model.inference.1}  
{\rm 
(Numenta inference algorithm)

$O$ is an object learned by Algorithm 
\ref{algorithm.subsect.model.learning.1}.
It is assumed that this object is observed
in this algorithm. 
This algorithm runs on each cortical column. 
The positive constants $\theta_p^{out},$ $\theta_b^{out},$     
$\theta_p^{in},$ $\theta_b^{in},$ and $\theta^{loc}$ 
are thresholds.  
In this algorithm, steps 
\ref{step.algorithm.subsect.model.inference.1.7} to  
\ref{step.algorithm.subsect.model.inference.1.14}
are repeated from the second round onwards.  
\begin{enumerate} 
\item 
\label{step.algorithm.subsect.model.inference.1.1}  
Set $t = 0,$ $A_t^{loc} = 0,$ 
$A_t^{in} = 0,$ 
$A_t^{out} = 0,$ and  
$\overline{A}_t^{out} = 0.$   
\item
\label{step.algorithm.subsect.model.inference.1.2} 
This step is stage $\bigcirc$\hspace*{-8pt}1\  
in Figure \ref{eq.subsect.hac_lpah.structure.1}.    
Select a location on $O$ at random.   
However, it is unknown which of the learned locations this location is. 
\item
\label{step.algorithm.subsect.model.inference.1.3}
This step is stage $\bigcirc$\hspace*{-8pt}3\ .  
Obtain $W^{in}_t$ from the sensory input at the location of 
step \ref{step.algorithm.subsect.model.inference.1.2}.
For every cell $c = (ij)$ in the sensory layer
(the $j$-th cell in the $i$-th mini-column), 
calculate the activity of $c$: 
\[
a_{ij,t}^{in} :=  
\left\{
\begin{array}{lll} 
1 & & \mbox{if}\ i \in W_t^{in} 
\\
[5pt] 
0 & & \mbox{otherwise}. 
\end{array}
\right.
\]
\item
\label{step.algorithm.subsect.model.inference.1.4} 
This step is stage $\bigcirc$\hspace*{-8pt}5\ .   
For every cell $k$ in the output layer,  
calculate 
\begin{equation}
a_{k,t}^{out} := 
\left\{
\begin{array}{lll} 
1 & & \mbox{if}\hspace{8pt}  
\displaystyle{\mathop{\sum}_{i,j}} f_{ijk} \cdot a_{ij,t}^{in}
\geq \theta_p^{out} 
\\
0 & & \mbox{otherwise}.  
\end{array}
\right.
\label{eq.step.algorithm.subsect.model.inference.1.4.2}  
\end{equation}
Thus, $A_t^{out} = \left(a_{k,t}^{out}\right)$ 
and the concatenated vector $\overline{A}_t^{out}$ are obtained. 
Set
\begin{equation}
W_t^{out} 
:= \left\{k:\ 
a_{k,t}^{out} = 1  
\right\}.
\label{eq.step.algorithm.subsect.model.inference.1.11.1}  
\end{equation}
\item 
\label{step.algorithm.subsect.model.inference.1.new5}  
This step is stage $\bigcirc$\hspace*{-8pt}6\ .
For every cell $k$ in the output layer,     
calculate 
\begin{equation}
\rho_{k,t}^{out} := 
\left\{
\begin{array}{ccl} 
1 & &\exists d:\ 
\overline{A}_{t}^{out} \iprod 
D_{kd}^{out} 
\geq \theta_b^{out} 
\\
[5pt] 
0 & & \mbox{otherwise}
\end{array}
\right.
\label{eq.algorithm.subsect.model.inference.1.12.1.t}  
\end{equation}
and
\[
a_{k,t}^{out} := 
\left\{
\begin{array}{lll} 
1 & & \mbox{if}\ k \in W_t^{out}\ \mbox{and}\ 
\rho_{k,t}^{out} = 1
\\
0 & & \mbox{otherwise}. 
\end{array}
\right.
\]
Thus, $A_t^{out}$ and $\overline{A}_t^{out}$ are updated.  
\item
\label{step.algorithm.subsect.model.inference.1.5}
This step is stage $\bigcirc$\hspace*{-8pt}7\ .    
For every cell $c = (ij)$ in the sensory layer, 
calculate 
\begin{equation}
\varpi_{ij,t}^{in} := 
\left\{
\begin{array}{ccl} 
1 & & \mbox{if}\hspace{8pt}  
\displaystyle{\mathop{\sum}_k} 
f_{ijk} \cdot a_{k,t}^{out}
\geq \theta_p^{in}
\\
0 & & \mbox{otherwise,} 
\end{array}
\right.
\label{eq.step.algorithm.subsect.model.inference.1.5.1}  
\end{equation} 
and update the activity of $c$:
\begin{equation}
a_{ij,t}^{in} := 
\left\{
\begin{array}{lll} 
1 & & \mbox{if}\ 
a_{ij,t}^{in} = 1 
\mbox{ and } \varpi_{ij,t}^{in} = 1 
\\
[5pt]
0 & & \mbox{otherwise}.  
\end{array}
\right.
\label{eq.step.algorithm.subsect.model.inference.1.5.2}  
\end{equation}
\item
\label{step.algorithm.subsect.model.inference.1.6}  
This step is stage $\bigcirc$\hspace*{-8pt}4\ .  
For every cell $\gamma$ in the location layer,
calculate 
\begin{equation}
\pi_{\gamma,t}^{loc} := 
\left\{
\begin{array}{ccl} 
1 & & \exists d:\ D_{\gamma,d}^{loc} 
\iprod A_{t}^{in} \geq \theta^{loc} 
\\
[5pt] 
0 & & \mbox{otherwise},  
\end{array}
\right.
\label{eq.step.algorithm.subsect.model.inference.1.14.1}  
\end{equation} 
where $A_{t}^{in} := \left(a_{ij,t}^{in}\right).$ 
Let $\vec{\phi}_{ih,t}$ be the location vector of the $h$-th cell 
in the module $i,$
and set 
\[
\Phi_{i,t}^{sense} := \left\{
\begin{array}{lll}
\left\{
\vec{\phi}_{ih,t}:\ \mbox{$\gamma = (ih)$ satisfies }
\pi_{\gamma,t}^{loc} = 1
\right\}
& &
\mbox{if } \exists \gamma = (ih):\ \pi_{\gamma,t}^{loc} = 1 
\\
[10pt]
\emptyset & & \mbox{otherwise}.    
\end{array}  
\right.
\]
Note that the elements of 
$\Phi_{i,t}^{sense}$ may indicate not only the true 
location on $O$ but also 
other locations on $O$ or 
locations on objects other than $O.$
For example, in Figure \ref{eq.sect.object.1.ABC} (A),
if the true location is the end point of the red vector, 
then this location and the locations of the all end points of the blue 
vectors in $O,$ $O',$ and $O''$ are indicated by 
$\Phi_{i,t}^{sense}.$    
\item
\label{step.algorithm.subsect.model.inference.1.7}  
Set $t := t + 1.$ 
\item
\label{step.algorithm.subsect.model.inference.1.8}  
This step is stage $\bigcirc$\hspace*{-8pt}1\ .  
Virtually or really, 
change the observation location on $O$ by (imaginary) motor input 
(see Remark \ref{remark.subsect.model.inference.3}).  
This motor input is represented  
by a movement vector $\vec{\delta}_{i,t}$ in each module $i$ 
of the location layer, and we obtain 
\begin{equation}
\Phi_{i,t}^{move} 
:= \left\{\vec{\phi}_t := \vec{\phi}_{t-1}  
+ \vec{\delta}_{i,t}:\ \vec{\phi}_{t-1} \in \Phi_{i,t-1}^{sense} 
\right\},
\label{eq.step.algorithm.subsect.model.inference.1.8.1}  
\end{equation}
where $\vec{\phi} + \vec{\delta}_{i,t}$ is considered on the torus made 
from the lattice of module $i.$ 
Make all cells corresponding to the elements of 
$\Phi_{i,t}^{move}$ active and the other cells inactive.  
Thus, $A_{t,move}^{loc} := A_t^{loc}$ is updated.  
\item
\label{step.algorithm.subsect.model.inference.1.9}  
This step is stage $\bigcirc$\hspace*{-8pt}2\ .
For every cell $c$ in the sensory layer, calculate 
\begin{equation}
\pi_{c,t}^{in} :=  
\left\{
\begin{array}{ccl} 
1 & &\exists d:\ D_{c,d}^{in} 
\iprod A_{t,move}^{loc} \geq \theta_b^{in} 
\\
[5pt] 
0 & & \mbox{otherwise}. 
\end{array}
\right.
\label{eq.algorithm.subsect.model.inference.1.pi}  
\end{equation}
Cell $c$ is predictive if and only if $\pi_{c,t}^{in} = 1.$   
\item
\label{step.algorithm.subsect.model.inference.1.10}  
This step is stage $\bigcirc$\hspace*{-8pt}3\ .
Move to a new observation location on $O$ 
by the movement vector $\vec{\delta}_{i,t}$ 
in step \ref{step.algorithm.subsect.model.inference.1.8}.  
Then, 
obtain $W^{in}_t$ from the sensory input at this location.  
For every cell $c = (ij)$ in the sensory layer, calculate 
the activity of $c$: 
\begin{equation}
a_{ij,t}^{in} := 
\left\{
\begin{array}{lll} 
1 & & \mbox{if}\ i \in W_t^{in}\ \mbox{and}\ \pi_{ij,t}^{in} = 1 
\\
[5pt]
1 & & \mbox{if}\ i \in W_t^{in}\ \mbox{and}\ \forall \mbox{ cell }k 
\in \mbox{mini-column $i$},   
\pi_{ik,t}^{in} = 0 
\\
[5pt] 
0 & & \mbox{otherwise}. 
\end{array}
\right.
\label{eq.step.algorithm.subsect.model.inference.1.10.1}  
\end{equation}
Thus, $A_t^{in}$ is updated. 
\item
\label{step.algorithm.subsect.model.inference.1.11}  
This step is stage $\bigcirc$\hspace*{-8pt}5\ .
For every cell $k$ in the output layer, calculate  
(\ref{eq.step.algorithm.subsect.model.inference.1.4.2}). 
Thus, $A_t^{out}$ and $\overline{A}_t^{out}$ are updated.  
Set $W_t^{out}$ using 
(\ref{eq.step.algorithm.subsect.model.inference.1.11.1}).  
\item
\label{step.algorithm.subsect.model.inference.1.12}  
This step is stage $\bigcirc$\hspace*{-8pt}6\ . 
For every cell $k$ in the output layer, calculate 
(\ref{eq.algorithm.subsect.model.inference.1.12.1.t})  
and the activity of $k$:
\begin{equation} 
a_{k,t}^{out} := 
\left\{
\begin{array}{lll} 
1 & & \mbox{if}\ k \in W_t^{out}\ \mbox{and}\ 
\rho_{k,t-1}^{out} = \rho_{k,t}^{out} = 1
\\
0 & & \mbox{otherwise}. 
\end{array}
\right.
\label{eq.algorithm.subsect.model.inference.1.12.2}  
\end{equation}
Thus, $A_t^{out}$ and $\overline{A}_t^{out}$ are updated.  

Stop this algorithm and we say that $O$ is recognized, 
if only the object $O$ is active in the sense of 
Definition \ref{definition.subsect.model.inference.1} 
described below. 
Otherwise, go to the next step.   
\item
\label{step.algorithm.subsect.model.inference.1.13}  
This step is stage $\bigcirc$\hspace*{-8pt}7\ . 
For every cell $c = (ij)$ in the sensory layer, calculate 
(\ref{eq.step.algorithm.subsect.model.inference.1.5.1}) and  
(\ref{eq.step.algorithm.subsect.model.inference.1.5.2}).   
Thus, $A_t^{in}$ is updated. 
\item
\label{step.algorithm.subsect.model.inference.1.14}  
This step is stage $\bigcirc$\hspace*{-8pt}4\ .
For each cell $\gamma = (ih)$ in the location layer, 
calculate (\ref{eq.step.algorithm.subsect.model.inference.1.14.1}),   
and set 
\begin{equation}   
\Phi_{i,t}^{sense}
:= \left\{
\begin{array}{lll}
\left\{
\vec{\phi}_{ih,t}:\ \mbox{$\gamma = (ih)$ satisfies }
\pi_{\gamma,t}^{loc} = 1
\right\}
& &
\mbox{if } \exists \gamma = (ih):\ \pi_{\gamma,t}^{loc} = 1 
\\
[10pt]
\Phi_{i,t}^{move} & & \mbox{otherwise}.    
\end{array}  
\right.
\label{eq.step.algorithm.subsect.model.inference.1.14.2}  
\end{equation}
Then, go to step \ref{step.algorithm.subsect.model.inference.1.7}.  
\end{enumerate}
}
\end{algorithm} 
\begin{remark} 
\label{remark.subsect.model.inference.2}  
\hspace{5pt} 
{\rm
Compared with the inference algorithms in \cite{HAC} and 
\cite{LPAH}, 
Algorithm \ref{algorithm.subsect.model.inference.1}  
has the following changes: 
\begin{itemize}
\item
The condition $\rho_{k,t}^{out} = 1$ in 
(\ref{eq.algorithm.subsect.model.inference.1.12.2})  
is added by the author.  
In \cite{HAC},  
the feedback stage $\bigcirc$\hspace*{-8pt}7\ , that is,  
the operation in 
steps \ref{step.algorithm.subsect.model.inference.1.5} and   
\ref{step.algorithm.subsect.model.inference.1.13}, 
is optional and definite formulae are omitted.
In the present study, as such formulae, 
(\ref{eq.step.algorithm.subsect.model.inference.1.5.1}) and      
(\ref{eq.step.algorithm.subsect.model.inference.1.5.2}) 
are added as matches to 
(\ref{eq.algorithm.subsect.model.inference.1.pi}) and   
(\ref{eq.step.algorithm.subsect.model.inference.1.10.1}), respectively.  
Condition $\rho_{k,t-1}^{out} = \rho_{k,t}^{out} = 1$ and 
the feedback stage $\bigcirc$\hspace*{-8pt}7 
\ result in the voting system described in \S 
\ref{subsect.model.activity}. 
\item
In Algorithm \ref{algorithm.subsect.model.inference.1},   
as in Algorithm \ref{algorithm.subsect.model.learning.1},  
cells are used instead of bumps 
to represent the activity in the location layer,
which differs from the algorithm in \cite{LPAH}.   
\end{itemize}
}
\end{remark}

\begin{remark} 
\label{remark.subsect.model.inference.3}  
\hspace{5pt} 
{\rm  
As policies for selecting motor inputs in step 
\ref{step.algorithm.subsect.model.inference.1.8},  
the paper \cite{CLH} lists model-based 
policies and model-free policies 
(see ``\textit{action policy}'' 
in \S 3.4 and \S 11 in \cite{CLH}). 
Model-based policies enable principled movement, such as moving a sensor 
to a location that will minimize the uncertainty 
of the currently observed object. 
In other words, the prediction can drive movement (cf. \S 1.7 of \cite{SB}). 
One must be able to compare the likelihoods  
of candidates for the observed object to achieve this minimization. 
The more candidates there exist, the costlier it becomes. 
However, avoiding this remains unclear to the author.  
Model-free policies are useful for purely sensory-based actions 
such as focusing on a prominent feature.}
\end{remark}

Algorithm 
\ref{algorithm.subsect.model.inference.1}  
can be considered as a Bayesian (or non-Bayesian) updating process,  
where object $O$ is the unknown parameter. 
The conditional probability is denoted 
by $P(\cdot |\cdot),$ a sensory input 
by $S_t,$ 
and a location on $O$ at time $t$ by $L_t(O).$  
As events, $S_t$ implies ``the sensory input is $S_t$''
and $L_t(O)$ implies ``the observation location is $L_t(O)$.''
Then, roughly speaking, the correspondences  
between the steps in Algorithm 
\ref{algorithm.subsect.model.inference.1}  
and probabilities calculated at each step are considered as 
summarized in 
Table \ref{table.subsect.model.inference.1}.  
\begin{table}[h] 
\caption{Correspondences  
between steps in Algorithm 
\ref{algorithm.subsect.model.inference.1}  
and probabilities} 
\begin{center}  
\begin{tabular}{c|c|c} 
steps \ref{step.algorithm.subsect.model.inference.1.9},
\ref{step.algorithm.subsect.model.inference.1.10}
& 
steps \ref{step.algorithm.subsect.model.inference.1.11} -- 
\ref{step.algorithm.subsect.model.inference.1.14}
& 
step \ref{step.algorithm.subsect.model.inference.1.8}
\\ \hline
$P(S_t|L_t(O))$ & $P(L_t(O)|S_t)$ & $P(L_{t+1}(O)|S_t)$  
\\ \hline
likelihood & posterior & prior     
\end{tabular}
\end{center}
\label{table.subsect.model.inference.1}  
\end{table}
These probabilities are based on the generative model that 
the cortical columns have, and 
$P(L_{t+1}(O)|S_t)$ essentially depends on the selection mechanism 
of the motor input 
as described in Remark 
\ref{remark.subsect.model.inference.3}.  
The details of Table \ref{table.subsect.model.inference.1},  
including the case of a  
non-Bayesian model, will be explained in \S \ref{sect.inference}. 

\subsection{Activity and selection of an object} 
\label{subsect.model.activity} 

On page 6 of \cite{HAC}, it is stated that 
``\textit{The set of active cells in the output layer represents the
objects that are recognized by the network. During inference
we say that the network unambiguously recognizes an object
when the representation of the output layer overlaps significantly
with the representation for correct object and not for any other
object.}''
Based on this concept, we define the following
(see step \ref{step.algorithm.subsect.model.inference.1.7}  
of Algorithm \ref{algorithm.subsect.model.inference.1}).
\begin{definition}
\label{definition.subsect.model.inference.1}  
\hspace{5pt} 
{\rm
Let $\theta_o^{out}$ and $\overline{\theta}_o^{out}$ 
be real numbers (thresholds)
such that $\theta_o^{out} > 0$ and $0 < \overline{\theta}_o^{out} < 1.$ 
Object $O$ is active in cortical column $C$
(more precisely, in the output layer of cortical column $C$)    
at time $t$ if   
\begin{eqnarray*}
& & 
\sharp\left(
\left\{k:\ k = 1 \mbox{ in $A_O^{out}$}\right\} \cap 
\left\{k:\ a_{k,t}^{out} = 1 \mbox{ in 
(\ref{eq.algorithm.subsect.model.inference.1.12.2})}\right\}
\right) 
\\
& \equiv & 
\sharp\left(\left\{k:\ k = 1 \mbox{ in $A_O^{out}$}\right\}
\cap 
\left\{k:\ a_{k,t}^{out} = 1 \mbox{ in 
(\ref{eq.step.algorithm.subsect.model.inference.1.11.1})}\right\}
\cap 
\left\{k:\ \rho_{k,t-1}^{out} = \rho_{k,t}^{out} = 1 
\right\}
\right) 
\geq \theta_o^{out}
\end{eqnarray*}
is satisfied in $C.$ 
Furthermore, object $O$ is active in the considered
cortical columns at time $t$ if 
\[
\sharp \left\{ \mbox{cortical column $C$}:\ 
\mbox{$O$ is active in $C$ at time $t$}\right\} \geq 
\overline{\theta}_o^{out} N^c 
\]
is satisfied. 
}
\end{definition} 
If Algorithm \ref{algorithm.subsect.model.inference.1} 
results in only one active object, $O,$ then 
$O$ is unambiguously recognized.  
In Algorithm \ref{algorithm.subsect.model.inference.1}, 
the flow of information 
between different cortical columns
is realized by 
(\ref{eq.algorithm.subsect.model.inference.1.12.1.t}) 
and 
(\ref{eq.algorithm.subsect.model.inference.1.12.2}).    
The other information flows remain in each cortical column.  
Condition $\rho_{k,t-1}^{out} = \rho_{k,t}^{out} = 1$  
causes and accelerates 
convergence of recognition, that is, 
convergence onto a representation for object $O.$  
This system 
is a \textit{voting} system among the cortical columns such that 
it combines the sensory inputs received by the cortical columns into 
a single perception (see \cite{HLKPA}, \cite{H2}, and \cite{CLH}). 

Throughout the duration of Algorithms \ref{algorithm.sect.related.1} 
and \ref{algorithm.sect.related.2}, 
a specific object must be continuously recognized. 
Additionally, in these algorithms,  
an object must be selected (randomly) from multiple active objects.
How should these processes be implemented ?
The author considers the following 
as one of the mechanisms for such an implementation:  
Let $C_1, C_2, \ldots, C_I$ be the cortical columns that store 
the considered objects 
$\mathcal{O} := \left\{O_1, O_2, \ldots, O_J\right\}.$ 
Each object is stored in one or more cortical columns.
Let $C$ be another cortical column with the following properties:
\begin{itemize} 
\item
Let $M_1, \ldots, M_m$ be the modules of the location layer of $C.$
Each object is represented by 
a set of cells $\{c_1, \ldots, c_m\}$ 
such that $c_j$ is selected from $M_j$ as follows:   
\begin{itemize} 
\item
For the object that is first recorded in $C,$ 
cell $c_j$ is randomly selected from $M_j.$
\item
Cell $c_j'$   
for the second and subsequent object recorded in $M_j$
is obtained by 
\[
\overrightarrow{c_j c_j'} = \vec{v}, 
\]
where
$c_j$ represents a recorded object $O \in \mathcal{O}$ and 
$\vec{v}$ is a movement vector. 
The object $O$ and the vector $v$ are 
common to $M_1,$ $\ldots,$ $M_m.$   
\end{itemize}
Connections exist between the cells representing an object 
$O \in \mathcal{O}$ 
in the output layers of  
$C_1, \ldots, C_I$ and the cells $\{c_1, \ldots, c_m\}$ 
representing $O$ in  
the location layer of $C.$
Thus, the set of cells $\{c_1, \ldots, c_m\}$ 
acts as a pointer to the cells representing $O$ 
in 
the output layers of $C_1,$ $\ldots,$ $C_I.$ 
For instance, as shown in Figures 4 and 5 in \cite{CLH}, 
$C_1,$ $\ldots,$ $C_I$ and $C$ form a hierarchical structure, 
and $C$ belongs to the layer one level above $C_1,$ $\ldots,$ $C_I.$
\item
A movement vector in the location layer of $C$ represents 
a movement from one object to another.
\item
The cells representing 
an object $O \in \mathcal{O}$ in the location layer of $C$ are connected 
to some cells in the sensory layer 
of $C.$ These cells represent the features of $O.$
\item
The output layer of $C$ is optional.
\end{itemize} 
We consider the following selection method for  
an object or objects 
from objects $O_1, \ldots, O_J$ in $C.$  
Only the selected object(s) in $C_1, \ldots, C_I$ and $C$ are activated. 
\begin{enumerate} 
\renewcommand{\labelenumi}{\theenumi}
\renewcommand{\theenumi}{(S\arabic{enumi})}
\item
\label{item.subsect.model.activity.S1}   
One of the cells representing objects stored 
in a module of the location layer of $C$ is randomly activated.
If the activated cell $c$ is one of the cells representing object $O,$ 
then the cells in the output layers of $C_1, ..., C_I$ connected to $c$ 
(the cells representing $O$) are also activated,    
and the cells representing $O$ 
in the location layer of $C$ 
are activated.
In this way, one object is randomly selected.
\item
\label{item.subsect.model.activity.S2}       
A movement vector in the location layer of $C$
starting from object $O$ selected in \ref{item.subsect.model.activity.S1}   
results in the selection of another object.
\item
\label{item.subsect.model.activity.S3}     
To select objects with desired features $F,$ first the cells representing 
$F$ in the sensory layer of $C$ are activated. 
Then, the objects in the location layer of $C$  
that are connected to these cells are activated.
\end{enumerate} 
Furthermore, 
if a specific object $O$ must be continuously recognized 
in $C_1, \ldots, C_I,$ it is realized by activating $O$ in $C$ 
continuously.
Each object $O$ stored in $C$ can be considered as an abstraction of 
the object stored in $C_1, \ldots, C_I.$
For instance, 
it may be assumed that $C$ is storing a language and $O$ is  
the name of the object.

In \ref{item.subsect.model.activity.S1},    
initially only one cell is activated.
This process is not robust to noise; 
however, if more than one cell is randomly selected, 
a high probability that more than one object will be selected exists.

\subsection{Values of thresholds} 
\label{subsect.model.thresholds} 

We assume that 
relationships between 
the values of thresholds 
$\theta_p^{out},$ $\theta_b^{out},$ $\theta_o^{out},$     
$\theta_b^{in},$ and $\theta^{loc}$ 
in Algorithm \ref{algorithm.subsect.model.inference.1} 
and the constants $\sharp \{c_{ij}\},$
$\natural A_O^{out},$ and $N^{loc}$ are  
as follows (cf. Remark \ref{remark.subsect.model.learning.3}),
where the values in $(\ )$ are used in \cite{HAC} 
and/or \cite{LPAH}:  
\begin{itemize} 
\item
$\theta_p^{out} (=3) \leq \bar{c} := \sharp \{c_{ij}\}$
$(= $ 5--8),
\item
$\theta_b^{out} (= 18) \leq \theta_o^{out} (= 30 
\leq 40) \leq \natural \overline{A}_O^{out},$ 
\item 
$\theta_b^{in} (=$ 6--8) $\leq N^{loc} (= 10),$ 
\item
$\theta^{loc} (=8) \leq \bar{c}.$ 
\end{itemize} 
Note that $\theta^{loc}$ is only used in \cite{LPAH} 
and $\bar{c}$ is only used in \cite{HAC}.   
In \cite{HAC}, some permanence values used during learning, 
and in \cite{LPAH}, bumps are used to represent activity in the 
location layer.
Although these are not used in the present study,
the values listed above are also consistent in this study. 
Thresholds $\overline{\theta}_o^{out}$
in Definition \ref{definition.subsect.model.inference.1} and  
$\theta_p^{in}$ in Algorithm 
\ref{algorithm.subsect.model.inference.1} 
are not used in \cite{HAC} and \cite{LPAH}.  
The value of $\overline{\theta}_o^{out}$ should not be excessively 
small because when some overlap exists between
the SDR of the observed object $O$
and that of another object $O',$ this $O'$ may also be active.
The value of $\theta_p^{in}$
is as follows.
We set $w := \sharp W_t^{in}$ and $a := \natural A_O^{out}.$ 
For any cell $c_{ij}$ in the sensory layer, 
define a random variable $X_{ij}$ as the number of cells 
in the output layer such that they are connected to  
$c_{ij}$ by Algorithm \ref{algorithm.subsect.model.learning.1}.   
If the connections are generated independently of each other, 
then the probability of $X_{ij} \geq \theta_p^{in}$ is independent of $(ij)$ and
is given by 
\[
P\left(X_{ij} \geq \theta_p^{in}\right) 
= \mathop{\sum}_{r=\theta_p^{in}}^a 
\left(\begin{array}{c} a \\ r \end{array}\right)
\left(\Frac{\bar{c}}{w}\right)^r \left(1-\Frac{\bar{c}}{w}\right)^{a-r}. 
\]
Then, a criterion for 
selecting the value of $\theta_p^{in}$
is that $P\left(X_{ij} \geq \theta_p^{in}\right) \geq p$ 
for the desired probability $p.$

\section{Finding similar objects} 
\label{sect.related} 

The set of learned and considered objects 
is denoted by $\Omega.$ 
In this section, two algorithms are proposed
to find objects in $\Omega$ 
similar to a given object $O \in \Omega,$ based on 
Algorithm \ref{algorithm.subsect.model.inference.1}. 
Although the setting for the proposed algorithms is restricted, 
the author believes that 
the case it covers is fundamental. 
Each object comprises a set of  
(location, feature) pairs as described in \S \ref{sect.object}. 
Therefore, $O$ and $O'$ are similar if and only if 
the (location, feature) pairs on $O$ and $O'$ are similar. 

Each sensory feature $f$ corresponds to 
the SDR $W_t^{in} = W^{in}(f)$ 
in the sensory layer. 
Let $\mathcal{F}$ 
be the set of all SDRs in the sensory layer.
We introduce a distance function $D$ on $\mathcal{F}$ such that 
$D(W^{in}(f),W^{in}(g))$ is small if and only if features $f$ and $g$ 
are similar. 
For instance, colors with similar wavelengths, such as blue and purple, 
are often considered as similar features. 
Then, we assume that the brain knows that these are similar, that is, 
$D(W^{in}(\mbox{blue}), W^{in}(\mbox{purple}))$ is small.  
Note that distance $D$ does not necessarily represent 
the physical distance on the sensory layer.
Let $d$ be a nonnegative number, and  
for $W \in \mathcal{F},$ define 
a neighborhood $\mathcal{N}_d(W)$ by 
\begin{equation}
\mathcal{N}_d(W) 
:= \left\{W' \in \mathcal{F}:\ 
D(W,W') \leq d \right\}.
\label{eq.sect.related.NW} 
\end{equation}
If $d = 0,$ then $\mathcal{N}_d(W) = \{W\}.$ 
We consider $\mathcal{N}_d(W)$ 
as a set of SDRs similar to $W.$    

Algorithms \ref{algorithm.sect.related.1} and \ref{algorithm.sect.related.2} 
to find similar objects 
are based on Algorithm \ref{algorithm.subsect.model.inference.1}.
The author believes 
that it would be preferable to make as few changes as possible 
from Algorithm \ref{algorithm.subsect.model.inference.1}.  
In Algorithm \ref{algorithm.sect.related.1},  
the only essential change is  
the replacement of $W_t^{in}$ with $\mathcal{N}_d(W_t^{in}),$ 
and several additional changes exist in 
Algorithm \ref{algorithm.sect.related.2}. 
If $d = 0,$
Algorithm \ref{algorithm.sect.related.1} 
is essentially the same as 
Algorithm \ref{algorithm.subsect.model.inference.1}. 
Algorithms \ref{algorithm.sect.related.1} 
and \ref{algorithm.sect.related.2} stop at a specified time. 
Algorithm \ref{algorithm.subsect.model.inference.1} 
can also find objects similar to the observed object $O,$ 
provided that it stops before converging on the representation of $O.$
Let $C_1, \ldots, C_I,$ and $C$ be the cortical columns described 
in \S \ref{subsect.model.activity},  
and let $O$ and $O'$ be objects stored in $C_1, \ldots, C_I.$  
The similarity relationship between $O$ and $O'$ obtained by these algorithms 
can be recorded as the positional relationship between $O$ and $O'$ 
in the location layer of $C$ by arranging or rearranging
similar objects close together.  
This is learning the ``similarity'' between objects.

\newpage
\begin{algorithm} 
\label{algorithm.sect.related.1} 
{\rm 
(Finding objects in $\Omega$ similar to the given object $O$)

$T$ is a nonnegative integer and $d$ is a nonnegative real number. 
Using Algorithm \ref{algorithm.subsect.model.inference.1}, 
the object $O$ has been recognized in the considered cortical columns.
It is assumed that 
$O$ can always be referred to (see \S \ref{subsect.model.activity}),
and all the real movement vectors and observation locations 
are on $O.$ 
\begin{enumerate} 
\item
\label{item.algorithm.sect.related.1.step.1} 
Set $t = 0,$ 
$A_t^{loc} = 0,$ 
$A_t^{in} = 0,$ 
$A_t^{out} = A_O^{out},$ and 
$\overline{A}_t^{out} = \overline{A}_O^{out}$ 
because $O$ has been already recognized. 
\item
\label{item.algorithm.sect.related.1.step.2} 
Select a location on $O$ at random.   
Then, this location is recognized. 
\item
\label{item.algorithm.sect.related.1.step.3} 
This step is the same as step 
\ref{step.algorithm.subsect.model.inference.1.3}
of Algorithm \ref{algorithm.subsect.model.inference.1}, 
except 
$W_t^{in}$ is replaced  
with $\mathcal{N}_d(W_t^{in}).$
\end{enumerate}
\begin{enumerate} 
\renewcommand{\labelenumi}{\theenumi}
\renewcommand{\theenumi}{\arabic{enumi}\hspace*{3pt}{}}
\setcounter{enumi}{3} 
\item
\hspace*{-3pt}-- 10. 
These steps are the same as steps 
\ref{step.algorithm.subsect.model.inference.1.4} to 
\ref{step.algorithm.subsect.model.inference.1.9} 
of Algorithm \ref{algorithm.subsect.model.inference.1}. 
\end{enumerate}
\begin{enumerate} 
\setcounter{enumi}{10} 
\item
\label{item.algorithm.sect.related.1.step.10}
This step is the same as step 
\ref{step.algorithm.subsect.model.inference.1.10}
of Algorithm \ref{algorithm.subsect.model.inference.1}, 
except 
$W_t^{in}$ is replaced  
with $\mathcal{N}_d(W_t^{in}).$
\item
This step is the same as step 
\ref{step.algorithm.subsect.model.inference.1.11}
of Algorithm \ref{algorithm.subsect.model.inference.1}. 
\item
\label{step.algorithm.sect.related.1.new13} 
For every cell $k$ in the output layer, calculate 
(\ref{eq.algorithm.subsect.model.inference.1.12.1.t})  
and (\ref{eq.algorithm.subsect.model.inference.1.12.2}).  
Thus, $A_t^{out}$ and $\overline{A}_t^{out}$ are updated.  

If $t = T,$ stop this algorithm.  
Otherwise, go to the next step.   

When this algorithm stopped, 
randomly select an object $O' \in \Omega$ such that it is active 
in the sense of 
Definition \ref{definition.subsect.model.inference.1}  
as an object similar to $O$
(see \S \ref{subsect.model.activity}).  
The end time $T$ may be determined dynamically. 
For example, when the number of active objects in the output layer
falls below a certain number, we set $t:=T.$
\end{enumerate}
\begin{enumerate} 
\renewcommand{\labelenumi}{\theenumi}
\renewcommand{\theenumi}{\arabic{enumi},}
\setcounter{enumi}{13} 
\item
15.  
These steps are the same as steps 
\ref{step.algorithm.subsect.model.inference.1.13} 
and \ref{step.algorithm.subsect.model.inference.1.14} 
of Algorithm \ref{algorithm.subsect.model.inference.1}. 
\end{enumerate} 
}
\end{algorithm}
Algorithm \ref{algorithm.sect.related.1}
uses all the mechanisms for 
convergence onto the representation of object(s) 
in Algorithm \ref{algorithm.subsect.model.inference.1},  
that is, the mechanisms of  
the algorithms in \cite{HAC} and \cite{LPAH}. 
Therefore, the convergence property of 
Algorithm \ref{algorithm.sect.related.1}
is essentially the same as that of  
Algorithm \ref{algorithm.subsect.model.inference.1}. 

According to \cite{D}, it is not possible for the brain to recognize 
multiple objects, simultaneously.   
Therefore, if the brain executes Algorithm \ref{algorithm.sect.related.1},
most of it (particularly 
the selection in step
\ref{step.algorithm.sect.related.1.new13})  
would be executed unconsciously.  

\begin{figure}[H]
\caption{Example of objects (they are the same objects as in Figure 
\ref{eq.sect.object.1.Obj})} 
\unitlength 1pt
\begin{center}
\begin{picture}(280,120)
\put(20,30){\line(1,0){40}}
\put(20,50){\line(1,0){60}}
\put(0,70){\line(1,0){80}}
\put(0,90){\line(1,0){80}}
\put(20,110){\line(1,0){20}}
\put(0,70){\line(0,1){20}}
\put(20,30){\line(0,1){80}}
\put(40,30){\line(0,1){80}}
\put(60,30){\line(0,1){60}}
\put(80,50){\line(0,1){40}}
\put(28,38){$\star$} 
\put(48,38){$\circ$} 
\put(27,57){$\Box$} 
\put(48,58){$\circ$} 
\put(68,58){$\circ$} 
\put(8,78){$\star$} 
\put(27,77){$\Box$} 
\put(47,77){$\Box$} 
\put(68,78){$\star$}
\put(28,98){$\star$}
\put(140,30){\line(1,0){20}}
\put(120,50){\line(1,0){60}}
\put(100,70){\line(1,0){80}}
\put(100,90){\line(1,0){80}}
\put(120,110){\line(1,0){20}}
\put(100,70){\line(0,1){20}}
\put(120,50){\line(0,1){60}}
\put(140,30){\line(0,1){80}}
\put(160,30){\line(0,1){60}}
\put(180,50){\line(0,1){40}}
\put(126,36){$E'$} 
\put(147,37){$\Box$} 
\put(128,58){$\circ$} 
\put(147,57){$\Box$} 
\put(167,57){$\Box$} 
\put(108,78){$\star$} 
\put(128,78){$\circ$} 
\put(148,78){$\circ$} 
\put(168,78){$\star$}
\put(128,98){$\star$} 
\put(220,30){\line(1,0){40}}
\put(220,50){\line(1,0){60}}
\put(200,70){\line(1,0){80}}
\put(200,90){\line(1,0){80}}
\put(260,110){\line(1,0){20}}
\put(200,70){\line(0,1){20}}
\put(220,30){\line(0,1){60}}
\put(240,30){\line(0,1){60}}
\put(260,30){\line(0,1){80}}
\put(280,50){\line(0,1){60}}
\put(228,38){$\bullet$} 
\put(248,38){$\circ$} 
\put(227,57){$\Box$} 
\put(248,58){$\circ$} 
\put(267,57){$\Box$} 
\put(208,78){$\bullet$} 
\put(227,77){$\circ$} 
\put(247,77){$\Box$} 
\put(268,78){$\bullet$}
\put(268,96){$\bullet$}
\put(40,10){$O$}  
\put(140,10){$O'$}
\put(240,10){$O''$}    
\end{picture}  
\end{center}  
\label{figure.sect.related.1} 
\end{figure}
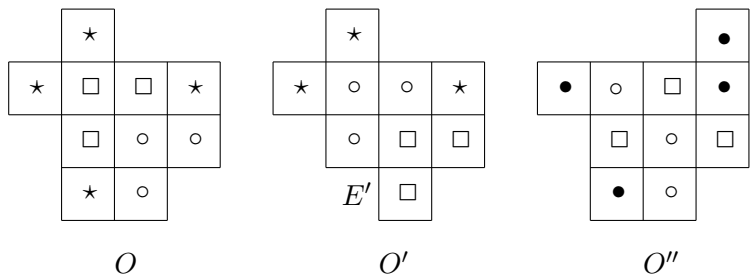
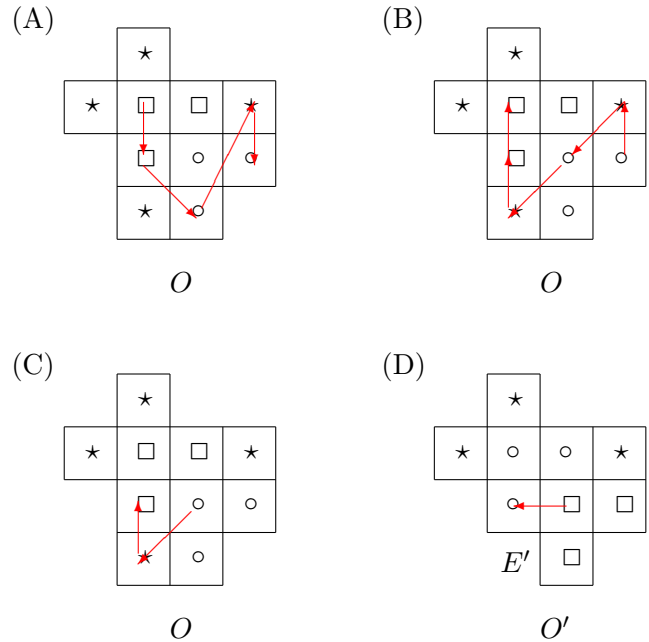
\begin{figure}[H]
\caption{Example of movement vectors} 
\unitlength 1pt
\begin{center}
\begin{picture}(250,130)
\put(0,110){(A)} 
\put(40,30){\line(1,0){40}}
\put(40,50){\line(1,0){60}}
\put(20,70){\line(1,0){80}}
\put(20,90){\line(1,0){80}}
\put(40,110){\line(1,0){20}}
\put(20,70){\line(0,1){20}}
\put(40,30){\line(0,1){80}}
\put(60,30){\line(0,1){80}}
\put(80,30){\line(0,1){60}}
\put(100,50){\line(0,1){40}}
\put(48,38){$\star$} 
\put(68,38){$\circ$} 
\put(47,57){$\Box$} 
\put(68,58){$\circ$} 
\put(88,58){$\circ$} 
\put(28,78){$\star$} 
\put(47,77){$\Box$} 
\put(67,77){$\Box$} 
\put(88,78){$\star$}
\put(48,98){$\star$}
\put(60,10){$O$}  
\put(50,82){\textcolor{red}{\vector(0,-1){20}}}
\put(50,58){\textcolor{red}{\vector(1,-1){20}}}
\put(72,42){\textcolor{red}{\vector(1,2){20}}}
\put(92,78){\textcolor{red}{\vector(0,-1){20}}}
\put(140,110){(B)} 
\put(180,30){\line(1,0){40}}
\put(180,50){\line(1,0){60}}
\put(160,70){\line(1,0){80}}
\put(160,90){\line(1,0){80}}
\put(180,110){\line(1,0){20}}
\put(160,70){\line(0,1){20}}
\put(180,30){\line(0,1){80}}
\put(200,30){\line(0,1){80}}
\put(220,30){\line(0,1){60}}
\put(240,50){\line(0,1){40}}
\put(188,38){$\star$} 
\put(208,38){$\circ$} 
\put(187,57){$\Box$} 
\put(208,58){$\circ$} 
\put(228,58){$\circ$} 
\put(168,78){$\star$} 
\put(187,77){$\Box$} 
\put(207,77){$\Box$} 
\put(228,78){$\star$}
\put(188,98){$\star$}
\put(200,10){$O$}  
\put(232,62){\textcolor{red}{\vector(0,1){20}}}
\put(232,82){\textcolor{red}{\vector(-1,-1){20}}}
\put(208,58){\textcolor{red}{\vector(-1,-1){20}}}
\put(188,42){\textcolor{red}{\vector(0,1){20}}}
\put(188,62){\textcolor{red}{\vector(0,1){20}}}
\end{picture}  
\begin{picture}(250,130)
\put(0,110){(C)} 
\put(40,30){\line(1,0){40}}
\put(40,50){\line(1,0){60}}
\put(20,70){\line(1,0){80}}
\put(20,90){\line(1,0){80}}
\put(40,110){\line(1,0){20}}
\put(20,70){\line(0,1){20}}
\put(40,30){\line(0,1){80}}
\put(60,30){\line(0,1){80}}
\put(80,30){\line(0,1){60}}
\put(100,50){\line(0,1){40}}
\put(48,38){$\star$} 
\put(68,38){$\circ$} 
\put(47,57){$\Box$} 
\put(68,58){$\circ$} 
\put(88,58){$\circ$} 
\put(28,78){$\star$} 
\put(47,77){$\Box$} 
\put(67,77){$\Box$} 
\put(88,78){$\star$}
\put(48,98){$\star$}
\put(60,10){$O$}  
\put(68,58){\textcolor{red}{\vector(-1,-1){20}}}
\put(48,42){\textcolor{red}{\vector(0,1){20}}}
\put(140,110){(D)} 
\put(200,30){\line(1,0){20}}
\put(180,50){\line(1,0){60}}
\put(160,70){\line(1,0){80}}
\put(160,90){\line(1,0){80}}
\put(180,110){\line(1,0){20}}
\put(160,70){\line(0,1){20}}
\put(180,50){\line(0,1){60}}
\put(200,30){\line(0,1){80}}
\put(220,30){\line(0,1){60}}
\put(240,50){\line(0,1){40}}
\put(208,37){$\Box$} 
\put(187,58){$\circ$} 
\put(208,57){$\Box$} 
\put(228,57){$\Box$} 
\put(168,78){$\star$} 
\put(187,78){$\circ$} 
\put(207,78){$\circ$} 
\put(228,78){$\star$}
\put(188,98){$\star$}
\put(185,35){$E'$}
\put(200,10){$O'$}  
\put(210,60){\textcolor{red}{\vector(-1,0){20}}}
\end{picture}  
\end{center} 
\label{figure.sect.related.ABC} 
\end{figure}

In the following, a broken line connecting the locations 
through which movement vectors pass is called a path.  
The starting point of a path is the starting point 
of the first movement vector, 
whereas the end point of a path is the end point of the last movement vector. 
In Algorithm \ref{algorithm.sect.related.1}, 
once an object becomes inactive, it cannot become active again. 
Assume that, in Figure \ref{figure.sect.related.1},  
features $\circ$ and $\Box$ are similar. 
If the path is as shown in 
Figure \ref{figure.sect.related.ABC} (A),
$O'$ is active and 
$O''$ is inactive from the third step onwards.
Next, assume that in Figure \ref{figure.sect.related.1}, 
features $\circ$ and $\Box$ are similar, 
and features $\star$ and $\bullet$ are similar. 
If the path is as shown in 
Figure \ref{figure.sect.related.ABC} (B),  
$O''$ is active and 
$O'$ is inactive from the third step onwards, 
because all corresponding paths on $O'$ are inactive 
(i.e., the third step locations of these paths do not exist or are inactive).
One of the reasons of this inactivity is that 
no location labeled $E'$ exists on $O'.$ 

As shown in the examples above, 
two cases exist in which an object becomes inactive:
\begin{enumerate} 
\renewcommand{\labelenumi}{\theenumi}
\renewcommand{\theenumi}{(N\Alph{enumi})}
\setcounter{enumi}{11}
\item
\label{item.sect.related.a}
No location exists on the object 
corresponding to the current location on $O.$ 
\setcounter{enumi}{5}
\item
\label{item.sect.related.b} 
Although corresponding locations exist, 
the features on none of these locations are similar
to the feature on the location on $O.$  
\end{enumerate} 
If the above inactivities are not acceptable, we can execute Algorithm 
\ref{algorithm.sect.related.2}, 
which is obtained by modifying 
Algorithm \ref{algorithm.sect.related.1}.
In this algorithm, if necessary, 
values of $\rho_{k,t}^{out}$s and $a_{k,t}^{out}$s 
are reset by 
(\ref{eq.item.algorithm.sect.related.1.step.14.2}) 
and an additional process is performed 
in step \ref{step.algorithm.subsect.model.inference.1.14}. 
Therefore, 
in either case \ref{item.sect.related.a} 
or \ref{item.sect.related.b}, 
the location information of the object is not lost, 
and the object continues to be observed in the next round.
\begin{algorithm} 
\label{algorithm.sect.related.2} 
{\rm 
(Finding objects in $\Omega$ similar to the given object $O$)

$T$ is a nonnegative integer and $d$ is a nonnegative real number. 
Using Algorithm \ref{algorithm.subsect.model.inference.1}, 
the object $O$ has been recognized in the considered cortical columns. 
It is assumed that 
$O$ can always be referred to, and 
all the real movement vectors and observation locations 
are on $O.$ 
For $O' \in \Omega,$ 
$\alpha_t(O')$ denotes a binary variable 
representing the activity of $O'$ 
in the sense of 
Definition \ref{definition.subsect.model.inference.1} 
at time $t,$ $1 \leq t \leq T.$  
$\Gamma$ is a positive integer such that $\Gamma \leq T.$
\begin{enumerate} 
\renewcommand{\labelenumi}{\theenumi}
\renewcommand{\theenumi}{\arabic{enumi},}
\item
2, 3. 
These steps are the same as steps  
\ref{item.algorithm.sect.related.1.step.1},
\ref{item.algorithm.sect.related.1.step.2},
and \ref{item.algorithm.sect.related.1.step.3}   
of Algorithm \ref{algorithm.sect.related.1}. 
\end{enumerate}
\begin{enumerate}  
\setcounter{enumi}{3} 
\item
\label{step.algorithm.sect.related.2.0}
This step is the same as step  
\ref{step.algorithm.subsect.model.inference.1.4}
of Algorithm \ref{algorithm.sect.related.1}. 
Set $\gamma_0(O') := 0$ for every active $O' \in \Omega.$ 
\end{enumerate}
\begin{enumerate} 
\renewcommand{\labelenumi}{\theenumi}
\renewcommand{\theenumi}{\arabic{enumi}\hspace*{3pt}{}}
\setcounter{enumi}{4} 
\item
\hspace*{-3pt}-- 12.   
These steps are the same as steps 
\ref{step.algorithm.subsect.model.inference.1.new5} 
to \ref{step.algorithm.subsect.model.inference.1.11} 
of Algorithm \ref{algorithm.sect.related.1}.
\end{enumerate}
\begin{enumerate} 
\setcounter{enumi}{12} 
\item
\label{step.algorithm.sect.related.2.1} 
For every cell $k$ in the output layer, calculate 
(\ref{eq.algorithm.subsect.model.inference.1.12.1.t})  
and (\ref{eq.algorithm.subsect.model.inference.1.12.2}).  
Thus, $A_t^{out}$ and $\overline{A}_t^{out}$ are updated.  
For every $O'$
that is active at time $t-1$ 
\begin{itemize}
\item  
set $\alpha_t(O') := 1$ if $O'$ is active, 
and set $\alpha_t(O') := 0$  
if $O'$ becomes inactive,
\item
calculate 
\[
\gamma_t(O') := \gamma_{t-1}(O') 
+ \left(1 - \alpha_t(O')\right).
\]
\end{itemize} 
If $\alpha_t(O') = 0,$
$\gamma_t(O') < \Gamma,$      
and $O'$ should be made active again, 
reset the activity of the cells in $\overline{A}_{O'}^{out}$ 
of each cortical column by resetting   
\begin{equation}
\rho_{k,t}^{out} := \rho_{k,t-1}^{out} 
\hspace{5pt}
\mbox{ and } \hspace{5pt} 
a_{k,t}^{out} := a_{k,t-1}^{out}
\label{eq.item.algorithm.sect.related.1.step.14.2}
\end{equation}  
for every cell $k$ in $\overline{A}_{O'}^{out}.$
Thus, $A_t^{out}$ and $\overline{A}_t^{out}$ are updated.  

If $t = T,$ stop this algorithm.  
Otherwise, go to the next step.   

When this algorithm stopped, 
randomly select an active object $O'$
as an object similar to $O$
(see \S \ref{subsect.model.activity}).   
The end time $T$ may be determined dynamically. 
\item
This step is the same as step
\ref{step.algorithm.subsect.model.inference.1.13} 
of Algorithm \ref{algorithm.sect.related.1}.  
\item
If there exists no $O'$
reactivated in 
step \ref{step.algorithm.sect.related.2.1},
this step is the same as step 
\ref{step.algorithm.subsect.model.inference.1.14} 
of Algorithm \ref{algorithm.sect.related.1}. 

If there exists such an $O',$
add the end point of every current path on every such $O'$ 
that satisfies \ref{item.sect.related.b} or 
\ref{item.sect.related.a} 
to $\Phi_{i,t}^{sense}$ in each cortical column. 
In case \ref{item.sect.related.a}, 
the set $\Phi_{i,t}^{sense}$ contains such end points  
as virtual position vectors. 

Then, go to step \ref{step.algorithm.subsect.model.inference.1.7}.  
\end{enumerate}
}
\end{algorithm}

\begin{remark} 
\label{remark.sect.related.1} 
\hspace{5pt} 
{\rm
Algorithm \ref{algorithm.sect.related.2} considers both 
\ref{item.sect.related.b} and \ref{item.sect.related.a}; 
however, it could also consider just one or the other.
For instance, to consider only \ref{item.sect.related.b}, 
Algorithm \ref{algorithm.sect.related.2} is changed as follows:
\begin{itemize}
\item
In step \ref{step.algorithm.sect.related.2.1}, 
reset the activity 
using (\ref{eq.item.algorithm.sect.related.1.step.14.2})
only if there exists a path on $O'$ such that it satisfies
\ref{item.sect.related.b}.
Therefore, if no path on $O'$ satisfies \ref{item.sect.related.b}, 
$O'$ is not reactivated.  
\item
In step \ref{step.algorithm.subsect.model.inference.1.14},  
perform processing only for \ref{item.sect.related.b}.   
\end{itemize}  
}
\end{remark}
\begin{remark} 
\label{remark.sect.related.3} 
\hspace{5pt} 
{\rm
In the \ref{item.sect.related.b} case,
step \ref{step.algorithm.subsect.model.inference.1.14}  
is implemented by rewriting step 
\ref{step.algorithm.subsect.model.inference.1.13} 
as follows:
\begin{enumerate} 
\renewcommand{\labelenumi}{\theenumi}
\renewcommand{\theenumi}{\arabic{enumi}'}
\setcounter{enumi}{13} 
\item
This step is the same as step
\ref{step.algorithm.subsect.model.inference.1.13} 
of Algorithm \ref{algorithm.sect.related.1},
except that for every object $O'$ reactivated in step
\ref{step.algorithm.sect.related.2.1}, 
change (\ref{eq.step.algorithm.subsect.model.inference.1.5.2})  
to
\[
a_{ij,t}^{in} := 
\left\{
\begin{array}{lll} 
1 & & \mbox{if}\ 
\pi_{ij,t}^{in} = 1 
\mbox{ and } \varpi_{ij,t}^{in} = 1 
\\
[5pt]
0 & & \mbox{otherwise}.  
\end{array}
\right.
\]
\end{enumerate}
Then, in step \ref{step.algorithm.subsect.model.inference.1.14},  
the end point of every current path on $O'$ 
that satisfies \ref{item.sect.related.b} is automatically added 
to $\Phi_{i,t}^{sense}.$ 
In the \ref{item.sect.related.a} case,
implementing step 
\ref{step.algorithm.subsect.model.inference.1.14}
would require a mechanism for information
to pass directly from the output layer to the location layer.
This requires making the model shown 
in Figure \ref{eq.subsect.hac_lpah.structure.1} 
more complicated. It is unclear to the author whether
Algorithm \ref{algorithm.sect.related.2} 
can be rewritten to avoid this complication.
}
\end{remark}  
In step \ref{step.algorithm.subsect.model.inference.1.12},
$\alpha_t(O') = 0$ implies that  
\[
\sharp\left(\left\{k:\ k = 1 \mbox{ in $A_{O'}^{out}$}\right\}
\cap 
\left\{k:\ \rho_{k,t-1}^{out} = 1 \right\}\right)
\geq \theta_o^{out}
\]
holds for $\overline{\theta}_o^{out} N^c$ or more cortical columns 
and
\begin{eqnarray*}
& & 
\sharp\left(\left\{k:\ k = 1 \mbox{ in $A_{O'}^{out}$}\right\}
\cap 
\left\{k:\ \rho_{k,t-1}^{out} = 1 \right\}\right.
\\
& & 
\left.
\hspace*{10pt} 
\cap 
\left\{k:\ \rho_{k,t}^{out} = 1 \right\}
\cap
\left\{k:\ a_{k,t}^{out} = 1 \mbox{ in 
(\ref{eq.step.algorithm.subsect.model.inference.1.11.1})}\right\}
\right) 
< \theta_o^{out}
\end{eqnarray*}
holds for $\left(1-\overline{\theta}_o^{out}\right) N^c$ 
or more cortical columns. 

We now provide some examples of how 
Algorithm \ref{algorithm.sect.related.2} runs.
Assume that 
$\Omega = \{O, O', O''\}$
as shown in Figure \ref{figure.sect.related.1} and 
features $\circ$ and $\Box$ are similar. 
As mentioned above,          
if the path is as shown in 
Figure \ref{figure.sect.related.ABC} (B) 
and Algorithm \ref{algorithm.sect.related.1} is used,  
two paths on $O'$ are active at the end of the second movement vector
and both 
paths become inactive at the end of the third movement vector. 
One of the end points is location $E'.$
Suppose $O'$ is to be reactivated.
Then, 
$O'$ becomes active again by 
(\ref{eq.item.algorithm.sect.related.1.step.14.2})
and end point $E'$ on one of the two paths above  
is recorded as the virtual end point of 
a vector in $\Phi_{i,t}^{sense}$ $(t = 3)$ by
step \ref{step.algorithm.subsect.model.inference.1.14}. 
In this step, 
the third and fourth movement vectors in 
Figure \ref{figure.sect.related.ABC} (B) 
on $O,$
that is, the movement vectors in 
Figure \ref{figure.sect.related.ABC} (C),   
are translated to a movement vector on $O'$ as in 
Figure \ref{figure.sect.related.ABC} (D).  
One of the two locations at time $t+1 = 4$ is the end point of this 
movement vector.  

Algorithm \ref{algorithm.sect.related.2} 
for $\Gamma = 1$ is nothing but 
Algorithm \ref{algorithm.sect.related.1}.  
As well as $O',$ 
suppose $O''$ should also be reactivated if it becomes inactive.
Assume that features $\circ$ and $\Box$ are similar.  
If $T = 5,$ $\Gamma = 3,$ 
and the path is as shown in 
Figure \ref{figure.sect.related.ABC} (B), 
then $\gamma_T(O') = 1 < \Gamma,$ $\gamma_T(O'') = 2 < \Gamma.$
Therefore, both $O'$ and $O''$ are active at time $T,$
and one of them is selected as an object similar to $O.$
If the value of $\Gamma$ is changed to 2,  
then $O'$ is active and $O''$ is inactive at time $T;$  
thus, $O'$ is selected. 
For location $L,$ the neighborhood $\mathcal{N}_d(L)$ 
in the location layer can be considered in the same manner as in 
(\ref{eq.sect.related.NW}), 
where $d$ is a value of a distance function 
(cf. \S 9.10 of \cite{CLH}).  
In this case, the ``corresponding location'' in 
\ref{item.sect.related.a} and \ref{item.sect.related.b} can be  
considered to be the ``location $L'$ nearest to the 
corresponding (virtual) location $L$ with $L' \in \mathcal{N}_d(L)$,'' 
and similarly for Algorithms   
\ref{algorithm.sect.related.1} and 
\ref{algorithm.sect.related.2}.  

For some simple cases, 
numerical experiments were conducted  
to obtain the probability of an object being active 
at the end of Algorithm  
\ref{algorithm.sect.related.2}.
It was 
assumed that no noise existed when 
Algorithm \ref{algorithm.sect.related.2} was executed. 
We considered a pair of observed object $O$ and another 
object $O' \in \Omega.$
In the experiments, 1000 randomly generated pairs of $(O,O')$ 
were used 
for each of the cases of $T = 3, 4, 5$ and $\Gamma = 1, 2.$ 
The experimental settings were as follows: 
\begin{itemize}
\item
Both objects $O$ and $O'$ 
comprise $5 \times 5$ grid locations. 
\item
The number of types of groups of similar features is 5.
These five types of features 
are uniformly randomly placed in 25 locations on 
each of objects $O$ and $O'.$ 
\item
Let $\overline{p}$ 
be a path connecting $T$ movement vectors on $O$ such that  
each path $\overline{p}$ for $T = 3, 4, 5$ 
is as depicted in Figure 
\ref{figure.sect.related.10}.   
These paths are used in the experiment.      
\item
$O'$ should be reactivated if it becomes inactive.
However,  
only movement vectors whose end points 
remain inside $O'$ are considered for simplicity.
Therefore, the operation for case 
\ref{item.sect.related.a} is not be executed. 
(See Remark \ref{remark.sect.related.1}.) 
\end{itemize} 
Let $N(\overline{p})$ 
be the number of paths with the same shape as $\overline{p}$ on 
$O'.$ 
Then, for $T = 3, 4,$ and $5,$ the values of $N(\overline{p})$ are  
40, 20, and 16, respectively. 
The results of experiments by using a Java program 
are as summarized in Table 
\ref{table.sect.related.1}. 
In this program, the Mersenne Twister
random number generator MTRandom.java was used (see \cite{PB}). 
The ``probability'' in this table refers to the probability 
(percent) 
that $O'$ is active 
at the end of Algorithm 
\ref{algorithm.sect.related.2}.   

\begin{figure}[H]
\caption{Paths connecting movement vectors} 
\unitlength 1pt
\begin{center}
\begin{picture}(340,80)
\put(0,30){\line(1,0){80}}
\put(0,30){\line(0,1){20}}
\put(0,50){\line(1,0){80}}
\put(20,30){\line(0,1){20}}
\put(40,30){\line(0,1){20}}
\put(60,30){\line(0,1){20}}
\put(80,30){\line(0,1){20}}
\put(10,40){\textcolor{red}{\vector(1,0){18}}}
\put(30,40){\textcolor{red}{\vector(1,0){18}}}
\put(50,40){\textcolor{red}{\vector(1,0){18}}}
\put(100,30){\line(1,0){100}}
\put(100,30){\line(0,1){20}}
\put(100,50){\line(1,0){100}}
\put(120,30){\line(0,1){20}}
\put(140,30){\line(0,1){20}}
\put(160,30){\line(0,1){20}}
\put(180,30){\line(0,1){20}}
\put(200,30){\line(0,1){20}}
\put(110,40){\textcolor{red}{\vector(1,0){18}}}
\put(130,40){\textcolor{red}{\vector(1,0){18}}}
\put(150,40){\textcolor{red}{\vector(1,0){18}}}
\put(170,40){\textcolor{red}{\vector(1,0){18}}}
\put(220,30){\line(1,0){100}}
\put(220,30){\line(0,1){20}}
\put(220,50){\line(1,0){100}}
\put(240,30){\line(0,1){20}}
\put(260,30){\line(0,1){20}}
\put(280,30){\line(0,1){20}}
\put(300,30){\line(0,1){40}}
\put(320,30){\line(0,1){40}}
\put(300,70){\line(1,0){20}}
\put(230,40){\textcolor{red}{\vector(1,0){18}}}
\put(250,40){\textcolor{red}{\vector(1,0){18}}}
\put(270,40){\textcolor{red}{\vector(1,0){18}}}
\put(290,40){\textcolor{red}{\vector(1,0){18}}}
\put(310,40){\textcolor{red}{\vector(0,1){18}}}
\put(30,10){$T = 3$} 
\put(140,10){$T = 4$} 
\put(260,10){$T = 5$} 
\end{picture}  
\end{center}  
\label{figure.sect.related.10} 
\end{figure}

\begin{table}[h]  
\caption{Probability (percent) 
that $O'$ is active at the end of Algorithm 
\ref{algorithm.sect.related.2}}
\begin{center}
\begin{tabular}{c||c|c|c|c|c|c}
$\Gamma$ 
& \multicolumn{3}{c|}{$\Gamma = 1$} 
& \multicolumn{3}{c}{$\Gamma = 2$} 
\\ \hline
$T$ 
& $T=3$ & $T=4$ & $T=5$ & 
$T=3$ & $T=4$ & $T=5$ 
\\ \hline
probability (\%)
& 7.0 & 0.4 & 0.1 & 59.6 & 10.2 & 1.6 
\end{tabular}
\end{center}
\label{table.sect.related.1} 
\end{table}

\begin{figure}[H]
\caption{Example of movement vectors} 
\unitlength 1pt
\begin{center}
\begin{picture}(390,130)
\put(0,110){(A)} 
\put(40,30){\line(1,0){40}}
\put(40,50){\line(1,0){60}}
\put(20,70){\line(1,0){80}}
\put(20,90){\line(1,0){80}}
\put(40,110){\line(1,0){20}}
\put(20,70){\line(0,1){20}}
\put(40,30){\line(0,1){80}}
\put(60,30){\line(0,1){80}}
\put(80,30){\line(0,1){60}}
\put(100,50){\line(0,1){40}}
\put(48,38){$\star$} 
\put(68,38){$\circ$} 
\put(47,57){$\Box$} 
\put(68,58){$\circ$} 
\put(88,58){$\circ$} 
\put(28,78){$\star$} 
\put(47,77){$\Box$} 
\put(67,77){$\Box$} 
\put(88,78){$\star$}
\put(48,98){$\star$}
\put(60,10){$O$}  
\put(48,78){\textcolor{red}{\vector(0,-1){20}}}
\put(48,58){\textcolor{red}{\vector(0,-1){20}}}
\put(52,38){\textcolor{red}{\vector(1,0){20}}}
\put(72,42){\textcolor{red}{\vector(1,1){20}}}
\put(92,64){\textcolor{red}{\vector(0,1){20}}}
\put(140,110){(B)} 
\put(200,30){\line(1,0){20}}
\put(180,50){\line(1,0){60}}
\put(160,70){\line(1,0){80}}
\put(160,90){\line(1,0){80}}
\put(180,110){\line(1,0){20}}
\put(160,70){\line(0,1){20}}
\put(180,50){\line(0,1){60}}
\put(200,30){\line(0,1){80}}
\put(220,30){\line(0,1){60}}
\put(240,50){\line(0,1){40}}
\put(208,37){$\Box$} 
\put(187,58){$\circ$} 
\put(208,57){$\Box$} 
\put(228,57){$\Box$} 
\put(168,78){$\star$} 
\put(187,78){$\circ$} 
\put(207,78){$\circ$} 
\put(228,78){$\star$}
\put(188,98){$\star$}
\put(185,35){$E'$}
\put(200,10){$O'$}  
\put(210,82){\textcolor{red}{\vector(-1,0){40}}}
\put(192,78){\textcolor{red}{\vector(1,0){40}}}
\put(188,58){\textcolor{red}{\vector(0,1){40}}}
\put(280,110){(C)} 
\put(340,30){\line(1,0){20}}
\put(320,50){\line(1,0){60}}
\put(300,70){\line(1,0){80}}
\put(300,90){\line(1,0){80}}
\put(320,110){\line(1,0){20}}
\put(300,70){\line(0,1){20}}
\put(320,50){\line(0,1){60}}
\put(340,30){\line(0,1){80}}
\put(360,30){\line(0,1){60}}
\put(380,50){\line(0,1){40}}
\put(348,37){$\Box$} 
\put(327,58){$\circ$} 
\put(348,57){$\Box$} 
\put(369,57){$\Box$} 
\put(308,78){$\star$} 
\put(327,78){$\circ$} 
\put(347,78){$\circ$} 
\put(368,78){$\star$}
\put(328,98){$\star$}
\put(325,35){$E'$}
\put(340,10){$O'$}  
\put(330,80){\textcolor{red}{\line(0,-1){40}}}
\put(330,40){\textcolor{red}{\line(1,0){20}}}
\put(350,40){\textcolor{red}{\line(1,1){20}}}
\put(370,60){\textcolor{red}{\vector(0,1){20}}}
\end{picture}  
\end{center} 
\label{figure.sect.related.O'} 
\end{figure}

\begin{remark} 
\label{remark.sect.related.2} 
\hspace{5pt} 
{\rm
We provide some supplementary explanations for 
Algorithms \ref{algorithm.sect.related.1}
and \ref{algorithm.sect.related.2}.   
\begin{itemize} 
\item
Assume that features $\circ$ and $\Box$ are similar, 
$\Gamma = 2,$ $T=5,$ and 
the path is as shown in 
Figure \ref{figure.sect.related.O'} (A).
Suppose $O'$ should be reactivated if it becomes inactive.
The three red paths in Figure \ref{figure.sect.related.O'} (B) 
are active at $t = 2$ and no active paths exist at $t = T.$  
In the red path in Figure \ref{figure.sect.related.O'} (C), 
there exists only one location such that it corresponds to either
\ref{item.sect.related.a} or \ref{item.sect.related.b}.
This path appears to be active even at $t = T$ when $\Gamma = 2.$
However, although the paths in Figure \ref{figure.sect.related.O'} (B) 
are active even at the second step, the path in
Figure \ref{figure.sect.related.O'} (C) 
becomes inactive at the second step. Therefore, the path in
Figure \ref{figure.sect.related.O'} (C) 
remains inactive after the second step.
We can rewrite Algorithm \ref{algorithm.sect.related.2}
to avoid this;
however, as it would be complicated, we do not proceed it.
\item
In Algorithms \ref{algorithm.sect.related.1}
and \ref{algorithm.sect.related.2},  
any object that was not activated in step 
\ref{step.algorithm.subsect.model.inference.1.4} 
will not be activated thereafter. 
If it is a problem, we may rerun Algorithms 
\ref{algorithm.sect.related.1} and \ref{algorithm.sect.related.2}.
It is easy if $T$ is a small value.  
\end{itemize} 
}
\end{remark}  

\section{Surprise and active inference} 
\label{sect.inference}

According to Friston's free-energy principle, 
active inference caused by surprise  
is considered (see \cite{F1}, \cite{F2}, 
and \cite{PPF}).
The surprise of sensory input $S_t$ is defined by 
$-\log P(S_t),$ the negative log evidence of $S_t.$ 
Thus, the smaller $P(S_t)$ is, the larger the surprise. 
In the present study, surprise also refers to sensory input that is 
(mostly) not predicted.
Thus, 
if in step \ref{step.algorithm.subsect.model.inference.1.10}  
of Algorithm \ref{algorithm.subsect.model.inference.1},   
most of mini-columns in $W_t^{in}$ satisfy the second condition 
of (\ref{eq.step.algorithm.subsect.model.inference.1.10.1}), 
then we consider that $S_t$ causing such a set $W_t^{in}$ is a surprise. 
As an active inference for surprise, we consider the following two 
essentially identical types:
\begin{enumerate} 
\renewcommand{\labelenumi}{\theenumi}
\renewcommand{\theenumi}{(\Roman{enumi})}
\item
\label{item.sect.inference.I}
Updating the prior to take 
the sensory input obtained at time $t$  
as the predicted sensory input at time $t+1.$   
An example of this active inference is 
when someone attempts to pour coffee into a cup from a pot 
but pours water instead, 
and then updates her/his knowledge about the contents of the pot. 
\item
\label{item.sect.inference.II}
Action at time $t+1$ to grasp the sensory input obtained at time $t$
as the predicted sensory input at time $t+1.$   
An example of this active inference is 
when someone sees something out of 
the corners of her/his eyes that is not predicted 
and turns her/his eyes in that direction. 
\end{enumerate}
We propose Algorithm \ref{algorithm.sect.inference.1}
as an algorithm such that it is a changed version of 
Algorithm \ref{algorithm.subsect.model.inference.1} 
to actively infer in both cases 
\ref{item.sect.inference.I} and \ref{item.sect.inference.II}. 
Condition $\rho_{k,t-1}^{out} = 1$ 
in (\ref{eq.algorithm.subsect.model.inference.1.12.2})  
interferes with this active inference.    
Therefore, in Algorithm \ref{algorithm.sect.inference.1}, 
the value of $\rho_{k,t-1}^{out}$ is set to 1 
in (\ref{eq.algorithm.sect.inference.1.rho})  
for the case where 
(\ref{eq.algorithm.subsect.bayes.stage3.1.1}) is satisfied. 

\begin{algorithm} 
\label{algorithm.sect.inference.1}
{\rm 
(Active inference for surprise)

The constants $\theta_w,$ $\theta'_w,$
and $\theta''_w$ are thresholds   
such that $0 < \theta_w < 1,$ $\theta'_w > 0,$ and 
$0 < \theta''_w < 1,$  
where $\theta_w$ is close to $1.$ 
\begin{enumerate} 
\renewcommand{\labelenumi}{\theenumi}
\renewcommand{\theenumi}{\arabic{enumi}\hspace{3pt}}
\item
\hspace*{-3pt}-- 10. 
These steps are the same as steps  
\ref{step.algorithm.subsect.model.inference.1.1} 
to \ref{step.algorithm.subsect.model.inference.1.9} 
of Algorithm \ref{algorithm.subsect.model.inference.1}. 
\end{enumerate} 
\begin{enumerate} 
\setcounter{enumi}{10}
\item
\label{step.algorithm.sect.inference.1.1}
This step is the same as step  
\ref{step.algorithm.subsect.model.inference.1.10}
of Algorithm \ref{algorithm.subsect.model.inference.1}, 
except
at the end of this step,  
check whether 
\begin{equation}
\sharp\left\{i \in W_t^{in}:\ \forall k 
\in \mbox{mini-column $i$}, 
\pi_{i,k,t}^{in} = 0 
\right\} \geq \theta_w \cdot \sharp W_t^{in} \geq \theta'_w  
\label{eq.algorithm.subsect.bayes.stage3.1.1.pre} 
\end{equation}
or not. 
\item
This step is the same as step  
\ref{step.algorithm.subsect.model.inference.1.11}
of Algorithm \ref{algorithm.subsect.model.inference.1}.  
\item
Check whether 
\begin{equation}
(\mbox{The number of cortical columns that satisfy 
(\ref{eq.algorithm.subsect.bayes.stage3.1.1.pre})}) 
\geq \theta''_w N^c, 
\label{eq.algorithm.subsect.bayes.stage3.1.1} 
\end{equation}
and perform the following. 
\end{enumerate} 
\textbf{The case where 
(\ref{eq.algorithm.subsect.bayes.stage3.1.1}) is not satisfied}. 
\begin{enumerate} 
\renewcommand{\labelenumi}{\theenumi}
\renewcommand{\theenumi}{\arabic{enumi},}
\setcounter{enumi}{12}
\item
14, 15. 
These steps are the same as steps  
\ref{step.algorithm.subsect.model.inference.1.12}, 
\ref{step.algorithm.subsect.model.inference.1.13},   
and \ref{step.algorithm.subsect.model.inference.1.14} 
of Algorithm \ref{algorithm.subsect.model.inference.1}. 
\end{enumerate} 
\textbf{The case where 
(\ref{eq.algorithm.subsect.bayes.stage3.1.1}) is satisfied}. 
\begin{enumerate} 
\setcounter{enumi}{12}
\item
\label{step.algorithm.sect.inference.1.3}
For every cell $k \in W_t^{out}$ in 
(\ref{eq.step.algorithm.subsect.model.inference.1.11.1}),    
set 
\begin{equation}
\rho_{k,t-1}^{out} = 1.
\label{eq.algorithm.sect.inference.1.rho}  
\end{equation}
For every cell $k$ in the output layer, calculate 
(\ref{eq.algorithm.subsect.model.inference.1.12.1.t})  
and (\ref{eq.algorithm.subsect.model.inference.1.12.2}).  
Thus, $A_t^{out}$ and $\overline{A}_t^{out}$ are updated.  

If there exist no active objects, stop this algorithm.
Then, this inference is a failure.
\item
This step is the same as step   
\ref{step.algorithm.subsect.model.inference.1.13} 
of Algorithm \ref{algorithm.subsect.model.inference.1}. 
\item
\label{item.algorithm.sect.inference.1.new.15}  
Calculate $\Phi_{i,t}^{sense}$ by 
(\ref{eq.step.algorithm.subsect.model.inference.1.14.2}).

Because (\ref{eq.algorithm.subsect.bayes.stage3.1.1}) is satisfied, 
probably this $\Phi_{i,t}^{sense}$ 
is very different from $\Phi_{i,t}^{move}.$ 

Then, go to step \ref{step.algorithm.subsect.model.inference.1.7}
described below.   
\setcounter{enumi}{7}
\item
Set $t := t + 1.$ 
\item
\label{item.algorithm.sect.inference.1.new.8}
Let the vector $\vec{\delta}_{i,t}$ in
(\ref{eq.step.algorithm.subsect.model.inference.1.8.1}) 
be $\vec{\delta}_{i,t} = 0.$
Then, the obtained locations correspond to the sensory input
$W_{t-1}^{in}.$ 
\item
This step is the same as step   
\ref{step.algorithm.subsect.model.inference.1.9} 
of Algorithm \ref{algorithm.subsect.model.inference.1}.

Then, go to step 
\ref{step.algorithm.subsect.model.inference.1.10} described above.  
\end{enumerate} 
}
\end{algorithm}

\begin{remark} 
\label{remark.sect.inference.2}
\hspace{5pt} 
{\rm
We provide some supplementary explanations for 
Algorithm \ref{algorithm.sect.inference.1}.
\begin{itemize} 
\item
If condition (\ref{eq.algorithm.subsect.bayes.stage3.1.1}) 
does not hold,
then Algorithm \ref{algorithm.sect.inference.1}
is the same as
Algorithm \ref{algorithm.subsect.model.inference.1}. 
In this case, although the activity in the second line of
(\ref{eq.step.algorithm.subsect.model.inference.1.10.1})  
may be reflected in
(\ref{eq.step.algorithm.subsect.model.inference.1.11.1}),   
its effect will usually disappear in
(\ref{eq.algorithm.subsect.model.inference.1.12.2}).  
\item
In step \ref{item.algorithm.sect.inference.1.new.8},
by setting $\vec{\delta}_{i,t} = 0,$
the active inference 
mapping $\Phi_{i,t-1}^{move} \rightarrow \Phi_{i,t}^{move}$  
is realized.
Because the selection of $\vec{\delta}_{i,t}$ takes no time,
this process is assumed to be instantaneous.  
\end{itemize} 
}
\end{remark}
\begin{remark} 
\label{remark.sect.inference.3}
\hspace{5pt} 
{\rm
Let the current time be $t.$         
The setting of the value of $\rho_{k,t-1}^{out}$
in (\ref{eq.algorithm.sect.inference.1.rho})     
contrasts with that of the value of $\rho_{k,t}^{out}$ in          
(\ref{eq.item.algorithm.sect.related.1.step.14.2}).
In (\ref{eq.algorithm.sect.inference.1.rho})     
the value of $\rho_{k,t-1}^{out}$ is reset,
whereas in (\ref{eq.item.algorithm.sect.related.1.step.14.2}) 
the value of $\rho_{k,t-1}^{out}$ is reused.
The set $\Phi_{i,t+1}^{move} = \Phi_{i,t}^{sense}$ 
induced by the reset (\ref{eq.algorithm.sect.inference.1.rho})     
causes a rewriting of $L_t(O)$ 
in Definition \ref{definition.sect.inference.1} below.
In this case, the prior probability obtained from the information
prior to time $t$ is lost,
and only the second term in 
(\ref{eq.definition.sect.inference.1.1}) 
remains.
}
\end{remark}
In the following, we consider Algorithm 
\ref{algorithm.sect.inference.1} 
from a non-Bayesian (modified Bayesian) inference perspective. 
The Bayesian updating process is defined as follows:  
\[
P_t(\theta|s_t) = \Frac{P(s_t|\theta)P_{t-1}(\theta|s_{t-1})}
{\sum_\theta P(s_t|\theta)P_{t-1}(\theta|s_{t-1})}
\hspace{10pt} 
(t = 1, 2, \ldots),
\]
where $\theta$ is the considered parameter, 
$(s_1, s_2, \ldots)$ are the given data, $P(\cdot |\cdot)$ 
is the conditional probability, and $P_0(\theta|s_0) := P(\theta).$ 
Thus, $P_{t-1}(\theta|s_{t-1})$ is the posterior to $\theta$ at time $t-1$ 
and the prior of $\theta$ at time $t.$
In \cite{HK}, 
the author considered the following   
non-Bayesian updating process:
\begin{equation}
P_t(\theta|s_t) = (1-\gamma_t)
\Frac{P(s_t|\theta)P_{t-1}(\theta|s_{t-1})}
{\sum_\theta P(s_t|\theta)P_{t-1}(\theta|s_{t-1})}
+ \gamma_t
\Frac{P(s_t|\theta)}{\sum_\theta P(s_t|\theta)}
\hspace{10pt} 
(t = 1, 2, \ldots),
\label{eq.sect.inference.1}
\end{equation}
where $0 \leq \gamma_t < 1.$ 
In \cite{HK}, 
this updating rule is interpreted as an overreacting to the observations.
The second term on the right-hand side of 
(\ref{eq.sect.inference.1})
can be nonzero even if the prior 
$P_{t-1}(\theta|s_{t-1})$ is zero.  
Thus, this term may be valid even if $s_{t-1}$ is not predicted, 
that is, $P(s_{t-1}|\theta) = 0.$ 
From this perspective, 
we consider the non-Bayesian formulation of  
Algorithm \ref{algorithm.sect.inference.1} as follows.  

Let $L_t(O)$ and $S_t$ be as defined in \S 
\ref{subsect.model.inference}, and $\Omega$ 
be as defined in \S \ref{sect.related}. 
Even if the brain is conscious of the movement vector, 
we consider $L_t(O)$ to be probabilistically determined.  
In the following, the inequality $P > 0$ means that  
$P$ is sufficiently large to be distinguished from $0.$ 
If not $P > 0,$ then we set $P = 0.$ 
We define  
\begin{equation}
\gamma_t = \gamma\left(S_t\right)
:= 
\left\{
\begin{array}{ccl}
1 & & \mbox{if (\ref{eq.algorithm.subsect.bayes.stage3.1.1})
is satisfied} 
\\
0 & & \mbox{if (\ref{eq.algorithm.subsect.bayes.stage3.1.1})
is not satisfied}
\end{array} 
\right. 
\label{eq.sect.inference.gamma}
\end{equation}
and 
\begin{eqnarray*}
\Omega_{t,0} & := &  
\left\{
O \in \Omega:\ P_t(L_{t+1}(O)|S_t) = 0
\right\}, 
\\
\Omega_{t,+} 
& := &  
\left\{
O \in \Omega:\ P_t(L_{t+1}(O)|S_t) > 0
\right\}, 
\\
\Omega_{t,L} & := & \left\{O \in \Omega:\ P(S_t|L_t(O)) > 0\right\}.
\end{eqnarray*}
Then, we define a non-Bayesian updating process.
It is also a type of state-space model. 
\begin{definition} 
\label{definition.sect.inference.1}
\hspace{5pt}
{\rm 
The stochastic process $L_t(O)$ is called 
a non-Bayesian updating process with 
discrete time $t = 1, 2, \ldots$ if it satisfies 
\begin{eqnarray}
P_t\left(L_t(O)|S_t\right) 
& = & \left(1-\gamma_t\right) 
\Frac{P(S_t|L_t(O)) \cdot P_{t-1}(L_t(O)|S_{t-1})}
{\sum_{O \in \Omega_{t,L} \cap \Omega_{t-1,+}} P(S_t|L_t(O)) 
\cdot P_{t-1}(L_t(O)|S_{t-1})}
\nonumber
\\
& & 
+ \gamma_t 
\Frac{P(S_t|L_t(O))}{\sum_{O \in \Omega_{t,L} \cap \Omega_{t-1,0}} P(S_t|L_t(O))}
\label{eq.definition.sect.inference.1.1}
\\
P_t\left(L_t(O)|S_t\right) 
& \rightarrow &  
P_t\left(L_{t+1}(O)|S_t\right).  
\label{eq.definition.sect.inference.1.2}
\end{eqnarray}
The posterior 
$P_t\left(L_t(O)|S_t\right)$ obtained by
(\ref{eq.definition.sect.inference.1.1}) 
is updated to the prior $P_t\left(L_{t+1}(O)|S_t\right)$ 
as (\ref{eq.definition.sect.inference.1.2})  
by
the movement vector in step 
\ref{step.algorithm.subsect.model.inference.1.8}  
of Algorithm \ref{algorithm.sect.inference.1} 
at time $t+1.$   
}
\end{definition}
For consistency between 
Definition \ref{definition.sect.inference.1}
and (\ref{eq.sect.inference.gamma}), 
we assume the following:
\begin{assumption}
\label{assumption.sect.inference.1}
\hspace{5pt}
{\rm
Assume that 
for every $O \in \Omega$ and $t,$  
if $\gamma_t = 1$ 
and $P(S_t|L_t(O)) > 0,$
then $P_{t-1}(L_t(O)|S_{t-1}) = 0.$  
}
\end{assumption}
From this assumption,  
we obtain that 
when $\gamma_t = 1,$ 
\[
P(S_t|L_t(O)) \cdot P_{t-1}(L_t(O)|S_{t-1}) = 0
\]
holds. 
This is consistent with the range of sum of 
the first term on the right-hand side of 
(\ref{eq.definition.sect.inference.1.1}).
It is natural to make Assumption \ref{assumption.sect.inference.1} 
when $\theta_w$ is sufficiently large.  
If $\theta_w$ is sufficiently large, 
then no predicted objects probably exist if $\gamma_t = 1.$ 
Assume $P_{t-1}(L_t(O)|S_{t-1}) > 0.$ 
Then, object $O$ was probably active at time $t-1.$ 
Therefore, if $P(S_t|L_t(O)) > 0,$ then we can believe that 
$S_t$ is predicted,
which implies $\gamma_t = 0.$   
Thus, we have Assumption \ref{assumption.sect.inference.1}. 

As described below, 
we consider Algorithm \ref{algorithm.sect.inference.1} 
to correspond to 
the non-Bayesian updating process
in Definition \ref{definition.sect.inference.1}.
This formulation 
is a different (non-)Bayesian type formulation from 
Friston's free-energy principle and it does not have the
universality that Friston's theory does.
However, it can make a description 
of surprises and active inferences 
such as those in \ref{item.sect.inference.I}  
and \ref{item.sect.inference.II}.   

We illustrate the probabilistic aspects 
of Algorithm \ref{algorithm.sect.inference.1}.
The likelihood $P(S_t|L_t(O))$ corresponds to
steps \ref{step.algorithm.subsect.model.inference.1.9} and 
\ref{step.algorithm.subsect.model.inference.1.10}.  
For instance, 
if we have no information regarding $P(S_t|L_t(O)),$
we consider
\begin{equation}
P(S_t|L_t(O))
=
\left\{
\begin{array}{lll}
\Frac{1}{\sharp \Omega_{t,L}} & & \mbox{if}\ O \in \Omega_{t,L} 
\\
0 & & \mbox{if}\ O \not\in \Omega_{t,L}.
\end{array}
\right.
\label{eq.sect.inference.5}
\end{equation}
The posterior $P_{t}(L_{t}(O)|S_{t})$ 
and prior $P_{t}(L_{t+1}(O)|S_{t})$ 
correspond to 
steps \ref{step.algorithm.subsect.model.inference.1.11} to 
\ref{step.algorithm.subsect.model.inference.1.14}
at time $t$ and step 
\ref{step.algorithm.subsect.model.inference.1.8} at time $t+1,$
respectively.
Depending on whether condition 
(\ref{eq.algorithm.subsect.bayes.stage3.1.1})  
holds, one of the terms on 
the right-hand side of 
(\ref{eq.definition.sect.inference.1.1}) is selected,  
and the probability $P_{t}(L_{t}(O)|S_{t})$ is obtained. 
Table \ref{table.subsect.model.inference.1} 
is based on this concept.

In both cases $\gamma_t = 0$ and $\gamma_t = 1,$ 
the posterior $P_t(L_t(O)|S_t)$ 
is updated by 
$L_t(O) \rightarrow L_{t+1}(O)$ 
in step
\ref{step.algorithm.subsect.model.inference.1.8}, 
$L_{t+1}(O) = L_t(O)$ when $\gamma_t = 1,$ 
and $P_t(L_{t+1}(O)|S_t)$ becomes the prior.  
In \cite{CLH},  
model-based and model-free 
policies are provided as policies for determining 
the motor input as described in Remark 
\ref{remark.subsect.model.inference.3}.  
In both learning and inference, both policies can work in concert. 
However, the policies in the inference following learned connections 
in Figure 
\ref{eq.subsect.hac_lpah.relation.connection.1}, that is, 
inference by Algorithm 
\ref{algorithm.subsect.model.inference.1}, will be primarily 
model-based. By contrast, the policies 
for the case $\gamma_t = 1$ and 
\ref{item.sect.inference.II} are considered to be primarily model-free. 
In step 
\ref{step.algorithm.subsect.model.inference.1.12} 
of Algorithm \ref{algorithm.sect.inference.1}  
for the case $\gamma_t = 1,$ 
the selection is assumed to be made uniformly random. 
Uniform randomness is also assumed in 
(\ref{eq.sect.inference.5}). 
(The random policy is a special case of the model-free policy.)
However, 
if the strength of the sensory input, $\sharp W_t,$  
differs for each object, 
this could be reflected in probabilities. 
Furthermore, using information from multiple cortical columns would 
yield more precise probabilities.

\section{Conclusion}
\label{sect.conclusion}

In this study, we studied the Numenta neocortex model. 
In \S \ref{sect.object} and \S \ref{sect.model}, 
the Numenta model was reviewed.   
Algorithms that find objects similar to the given object $O$
(Algorithms \ref{algorithm.sect.related.1} and 
\ref{algorithm.sect.related.2}) and 
an algorithm that actively infers surprise  
(Algorithm \ref{algorithm.sect.inference.1}) were proposed 
by slightly changing the Numenta inference model
(Algorithm \ref{algorithm.subsect.model.inference.1}).   

An important aspect of these algorithms is how 
the motor input (movement vector) is selected. 
As described in Remark \ref{remark.subsect.model.inference.3}  
and \S \ref{sect.inference}, 
model-based and model-free policies
to determine the motor input 
were proposed by \cite{CLH}. 
These policies enable the model to quickly identify the observed object 
and react to the surprise.
According to \cite{D}, \cite{F1}, \cite{F2}, and \cite{PPF}, 
the motor input appears to be generated unconsciously in numerous cases, 
and 
according to \cite{F1}, \cite{F2}, and \cite{PPF}, 
unconscious processing is generally performed to minimize 
the variational free energy. 
The policies proposed by \cite{CLH} and Algorithm 
\ref{algorithm.subsect.model.inference.1}
could be deemed to be a method of performing this minimization.
However, it seems that implementing conscious selection 
of the motor input remains unclear.
This is an open problem for the author.  

From the perspective of ``association,'' the setting for 
the algorithms in \S \ref{sect.related} is significantly limited.
To improve this, it is expected that research on association,
such as that described in Chapter 6 of \cite{H2} and 
\cite{VK}, which investigates association in
feedforward artificial neural networks, is useful.
In Algorithms \ref{algorithm.sect.related.1} 
and \ref{algorithm.sect.related.2}, 
the selection method of the motor input is important. 
This selection in the inference depends on the learning results. 
Thus, the learning method of the 
associations between objects is important. 
These are also open problems for the author.

\section*{Acknowledgements} 
This work was supported by JSPS KAKENHI Grant
Number JP22K11916.

\end{document}